\documentclass[preprint,secnumarabic,amssymb, nobibnotes, aps, prd]{revtex4-1}

\usepackage{bm}
\usepackage{amsmath}
\usepackage{graphicx}
\usepackage{booktabs}
\usepackage{float}
\usepackage{subfigure}
\usepackage{amssymb}
\usepackage{appendix}

\begin{document}

\title{A phase-field model for large-density-ratio two-phase flows based on discrete unified gas-kinetic scheme}%
\author{Zeren Yang}%
\email[Z. R. Yang: ]{zeren@mail.nwpu.edu.cn}
\author{Chengwen Zhong}
\email[Corresponding author: ]{zhongcw@nwpu.edu.cn}
\author{Congshan Zhuo}
\email[C. S. Zhuo: ]{zhuocs@nwpu.edu.cn}

\affiliation{National Key Laboratory of Science and Technology on Aerodynamic Design and Research, Northwestern Polytechnical University, Xi'an, Shaanxi 710072, China.}
\date{\today}%
\begin{abstract}
In this paper, a phase-field based model under the framework of discrete unified gas-kinetic scheme (DUGKS) for incompressible multiphase fluid flows is proposed. Two kinetic models are constructed to solve the conservative Allen-Cahn (A-C) equation that accounts for the interface behavior and the incompressible hydrodynamic equations that govern the flow field, respectively. With a truncated equilibrium distribution function as well as a temporal derivative added to the source term, the macroscopic governing equations can be exactly recovered from the kinetic models through the Chapmann-Enskog analysis. Calculation of source terms involving high-order derivatives existed in the quasi-incompressible model is simplified. A series of benchmark cases including four interface-capturing tests and four binary flow tests are carried out. Results compared to that of lattice Boltzmann method (LBM) have been obtained. A convergence rate of second-order can be guaranteed in the test of interface diagonal translation. The capability of present method in interface tracking that undergoes a severe deformation has been verified. Stationary bubble and spinodal decomposition problems, both with a density ratio as high as 1000, are conducted and reliable solutions have been provided. The layered Poiseuille flow with a large viscosity ratio is simulated and numerical results agree well with the analytical solutions. Variation of positions of the bubble front and spike tip during the evolution of Rayleigh-Taylor instability (RTI) has been predicted precisely. However, the detailed depiction of complicated interface patterns appeared in the evolution process is failed, which is mainly caused by the relatively large numerical dissipation of DUGKS compared to that of LBM. A high-order DUGKS is needed to overcome this problem.
\end{abstract}

\maketitle

\section{Introduction}
Numerical simulation of multiphase fluid flows have drawn attention of many researchers for decades due to its importance in scientific and engineering applications. With the rapid progress in computational technology, various numerical methods including the volume of fluid (VOF) method\cite{HIRT1981201}, level set approach\cite{SUSSMAN1994146}, front tracking method\cite{UNVERDI199225}, diffuse interface method\cite{Anderson1998,YUE2004} and smoothed particle hydrodynamics method\cite{Gingold1977Smoothed,Lucy1977A} have been developed. Among those methods, diffuse interface model has shown great advantage by virtue of its energy-based variational formalism as well as its simplification in the description of interface evolution. Many of kinetic schemes\cite{He2002,Gan2015Discrete,Wang2015Multiphase,pan_cheng_wang_xu_2017}, aiming at modeling phase behaviors at the mesoscopic level and bridging the gap between the macroscopic features and microscopic intermolecular interactions in multiphase systems, can be categorized into the diffuse interface method.

As one of many popular kinetic schemes, the lattice Boltzmann (LB) method have received great attention due to its simplicity in the application of intermolecular interactions. In the framework of LB method, mainly four kinds of multiphase models, including color-gradient model\cite{Gunstensen1991}, pseudopotential model\cite{Shan1993,Shan1994}, free-energy model\cite{Swift1995,INAMURO2004628}, and phase-field based model\cite{HE1999642,INAMURO2004628} are proposed based on different physical pictures. These LB models have been improved continuously by plenty of research since the moment they were born and significant progresses on their performance have been made\cite{LIU2016873,LIU2017337,CHEN20113577,Wen2017,Shao2014,Liang2014Phase}. While they share the same advantages of LB method, such as simplicity and efficiency, relatively low dissipation and intrinsic kinetic nature\cite{Aidun2010}, they are also limited by the drawbacks of LB method,
one of which is the required uniformity of the lattice structure imposed by the symmetry of the predefined lattice velocities. In order to capture the interface clearly and precisely, either high resolution scheme or adaptive mesh refinement (AMR) technique is usually employed in the simulation of multiphase problems. The usage of uniform mesh is surely a waste of time and computational resources, however, application of ARM technique to LB method would result in a loss of its simplicity. The multiphase lattice Boltzmann flux solver (MLBFS), aiming to combines the advantages of Navier-Stokes solvers and LB method in the simulation of multiphase flows, was first proposed by Wang and Shu\cite{Wang2015Multiphase,WANG2015336}. In their work, a fifth-order upwind scheme is adopted to solve the Cahn-Hilliard (C-H) equation that governs the evolution of interface and a lattice Boltzmann flux scheme is used to evaluate the flux at cell interface for the mass and momentum equation. A density ratio of 1000 with Reynolds number up to 3000 is achieved in the simulation of Rayleigh-Taylor instability (RTI). Pan\cite{pan_cheng_wang_xu_2017} developed a two-stage fourth-order gas-kinetic scheme (GKS) for the simulation of compressible multi-component flows. Based on a simplified two-species BGK model, a set of coupled Euler equations that accounts for different components are constructed and solved by a fourth-order gas-kinetic scheme. Various numerical tests including shock-bubble interaction, Rayleigh-Taylor instability and etc have verified the reliability of this approach. MLBFS and GKS show a common philosophy in the construction of flux since kinetic schemes, although different in detail, are introduced for both methods, which is also the only point where they have a relation with kinetic schemes. Furthermore, both of there two methods lack the ability of depicting the non-equilibrium effects in the flow field. Gan and Xu\cite{Gan2015Discrete} proposed a discrete Boltzmann model (DBM) to study the process of phase separation based on the work of Gonnella, Lamura and Sofonea\cite{Gonnella2007}. The interplay between interparticle force that drives changes and gradient force that opposes them is simulated and the non-equilibrium effects behind the phenomenon is investigated thoroughly, which provides a better understanding of the non-equilibrium behaviors underneath the phase separation process. Later, the DBM is applied to the simulation of Rayleigh-Taylor instability in compressible flows\cite{Lai2016Nonequilibrium,Lin2017Discrete}. The relations between effects of compressibility and global non-equilibrium intensity are investigated and a negative correlation is observed.The DBM provides an distinctive way for the explanation of physical phenomenon from a viewpoint of non-equilibrium effects. However, it suffers from the constraint on collision time and time step, which imposes a negative impact on its efficiency.

The discrete unified gas-kinetic scheme (DUGKS) proposed by Guo et al.\cite{Guo2013Discrete,Guo2015Discrete} is a new type of kinetic scheme which combines the advantages of both LB method in its discrete conservative collision operator and gas-kinetic scheme in its flux modeling\cite{Wang2016Comparison}. As a finite volume method, it overcomes the disadvantage of uniform mesh with which LB method has to comply and simplifies the calculating routine in the evaluation of flux at the cell interface. The fully coupling of streaming and collision processes in DUGKS ensures a low numerical dissipation feature. The semi-implicit treatment of the collision term in DUGKS also makes an improvement in its numerical stability\cite{Wang2015A}. Furthermore, the evolution equation rather than direct interpolation is employed in the evaluation of flux, which contributes to its asymptotic preserving (AP) feature\cite{Guo2015Discrete}. That means the time step in only limited by Courant-Friedrichs-Lewy (CFL) condition rather than the collision time in the framework of DUGKS. Compared to the other discrete velocity methods, DUGKS has shown a better performance in terms of modeling accuracy and computational efficiency when the hydrodynamic
flow regime is dominant\cite{WANG201833}. Due to its various advantages, DUGKS has been applied to modeling non-equilibrium flow\cite{Zhu2017,Zhu2017Numerical}, phonon transportation\cite{PhysRevE.96.063311}, binary gas mixtures\cite{PhysRevE.97.053306}, fluid-particle flows\cite{Tao2018,Huo2018}. Recently, Zhang and Guo\cite{Zhang2018} extend the DUGKS to two-phase flows based on a quasi-incompressible phase-field governing equation. The accuracy and stability of this model have been verified. However, the density contrast of different phases in their work is relatively low and no interface capturing test is given to evaluate the capability of DUGKS under such a circumstance. To make a further extension of DUGKS on the simulation of two-phase problems, we proposed a new model by applying the DUGKS to the solution of the conservative Allen-Cahn (A-C) equation\cite{Sun2007Sharp,Chiu2011,Geier2015Conservative}. The incompressible hydrodynamic equation by Liang\cite{Liang2014Phase} is also implemented in the DUGKS framework. Various interface capturing tests are conducted and binary flow cases with a high density ratio are studied.

The rest of this article is arranged as follows. In Sec. \ref{sec:METHODOLOGY}, the methodology of the proposed model for two-phase flows will be introduced. In Sec. \ref{SEC:INTERFACE-CAPTURING TESTS}, several benchmark tests are conducted to validate the capability of current method in capturing interface. In Sec. \ref{SEC:BINARYFLOWTESTS}, typical two-phase flow cases are carried out to verify the performance of our model. A brief summary is drawn in Sec. \ref{sec:CONCLUSION}.

\section{METHODOLOGY}\label{sec:METHODOLOGY}
\subsection{Governing equation for two-phase flows}
Based on the phase-field theory, a Helmholtz free energy functional dependent upon an order parameter $\phi$ is used to describe the thermodynamic behaviour of a two-phase fluid system\cite{Li2016Lattice},
\begin{equation}
F(\phi)=\int_{V}\bigg(\epsilon(\phi)+\frac{\kappa}{2}\vert{\nabla\phi}\vert^2\bigg)dV,
\end{equation}
where $V$ is the domain of the system, and $\epsilon(\phi)$ refers to the bulk energy density.
For binary fluids system, the function of bulk energy density usually has the following double-well form,
\begin{equation}
\epsilon(\phi)=\beta(\phi-\phi_H)^2(\phi-\phi_L)^2,
\end{equation}
which has two minima corresponding to the two phases of the fluid. The parameters $\beta$ and $\kappa$ are two positive constants determined by both the coefficient of surface tension $\sigma$ and the width of interface $W$,
\begin{equation}
\kappa=\frac{3}{2}\sigma{W},\ \beta=\frac{12\sigma}{W}.
\end{equation}
The chemical potential $\mu_\phi$ is defined as the variation of the free energy with respect to the order parameter,
\begin{equation}
\mu_\phi=\frac{\delta{F(\phi)}}{\delta\phi}=4\beta(\phi-\phi_H)(\phi-\phi_L)(\phi-\frac{\phi_H+\phi_L}{2})-\kappa\Delta\phi.
\end{equation}
The equation used for interface-tracking in present study is the following conservative Allen-Cahn equation\cite{Geier2015Conservative},
\begin{equation}
\frac{\partial\phi}{\partial{t}}+\nabla\cdot(\phi\bm{u})=\nabla\cdot[M_\phi(\nabla\phi-\theta\bm{n})],
\label{Eq:interfaceEq}
\end{equation}
where $t$ is the time, $\bm{u}$ is the transportation velocity, $M_\phi$ is the mobility coefficient and $\bm{n}$ is the local unit vector normal to the interface. $\theta$ is interpreted as a function of $\phi$,
\begin{equation}
\theta = \frac{-4(\phi-\phi_H)(\phi-\phi_L)}{W(\phi_H-\phi_L)}.
\end{equation}
The equilibrium profile of $\phi$ along the direction normal to the interface assumes a hyperbolic tangent form,
\begin{equation}
\phi(z) = \frac{\phi_H + \phi_L}{2} + \frac{\phi_H - \phi_L}{2}\text{tanh}\Big(\frac{2z}{W}\Big),
\end{equation}
where $z$ is the coordinate along interface normal. The location of actual interface is determined by $\phi = 0.5(\phi_H+\phi_L)$, where $\phi_H = 1$ denotes the heavy fluid and $\phi_L = 0$ indicates the light fluid.

For an exhaustive derivation of conservative ACE, one is recommended to refer to Ren and Song\cite{Ren2016Improved}. It is worth noting that divergence-free velocity condition was introduced during this derivation.

The hydrodynamic equations used for a two-phase fluid system are chosen to be
\begin{equation}\label{Eq:Eq08}
\nabla\cdot\bm{u} = 0,
\end{equation}
\begin{equation}
\frac{\partial(\rho\bm{u})}{\partial{t}}+\nabla\cdot(\rho\bm{u}\bm{u})=-\nabla{p} + \nabla\cdot[\mu(\nabla{\bm{u}}+\nabla{\bm{u}^T})] +\bm{F},
\label{Eq:HydrodynamicMacro}
\end{equation}
where $\bm{F}$ consists of the surface tension force $\bm{F}_s=\mu_\phi\nabla\phi$ and the gravitational force $\bm{G}$, if present.
The relationship between order parameter $\phi$ and density $\rho$ is
\begin{equation}
\rho = \frac{\rho_H-\rho_L}{\phi_H-\phi_L}(\phi-\phi_L)+\rho_L.
\label{Eq:rho}
\end{equation}
Substituting Eq.(\ref{Eq:interfaceEq}) and Eq.(\ref{Eq:Eq08}) into Eq.(\ref{Eq:rho}), we can get the non-conservative mass equation,
\begin{equation}
\frac{\partial\rho}{\partial{t}}+\nabla\cdot(\rho\bm{u})=\frac{\rho_H-\rho_L}{\phi_H-\phi_L}\nabla\cdot[M(\nabla\phi-\theta\bm{n})].
\end{equation}
The point we are trying to clarify here is that with the divergence-free velocity condition, the uniform conservative mass equation cannot be derived from the conservative Allen-Cahn equation. The conservativeness of mass equation can be guaranteed only if the density gradient equals zero, which means that mass generation or consumption exists during the process of phase transition when density contrast exists. As is depicted by Li\cite{Li2012Additional}, the uniform mass conservation and the incompressibility condition cannot be satisfied at the same time because of the volume diffusive flux across the interfacial region. Hence, it is not the mass parameter $\rho$ but the order parameter $\phi$ that conservativeness qualifies. One way to eliminate the non-conservative property of mass is to absorb the source term on the right hand side of mass equation into the velocity divergence. A novel model based on this idea has been proposed by Yang and Guo\cite{Yang2016Lattice}.

\subsection{\label{sec:subsecDUGSKforTwoPhase}DUGKS for two-phase Equations}
The discrete kinetic equations used to interpret phase equation Eq. (\ref{Eq:interfaceEq}) and hydrodynamic equation (\ref{Eq:HydrodynamicMacro}) are
\begin{equation}
\frac{\partial{f_i}}{\partial{t}}+\xi_i\cdot\nabla{f_i}=-\frac{f_i-f_i^{eq}}{\tau_f}+S^f_i,
\label{fdke}
\end{equation}
\begin{equation}
\frac{\partial{g_i}}{\partial{t}}+\xi_i\cdot\nabla{g_i}=-\frac{g_i-g_i^{eq}}{\tau_g}+S^g_i,
\label{gdke}
\end{equation}
where $f_i$ and $g_i$, corresponding to the phase order $\phi$ and density $\rho$, are the particle distribution functions in terms of position $\bm{x}$, discrete particle velocity $\bm{\xi}_i$ and time $t$. $\tau_f$ and $\tau_g$ are the relaxation times related to the mobility coefficient and dynamic viscosity, respectively. $f^{eq}_i$ and $g^{eq}_i$ are the equilibrium distribution functions with specific forms. $S^f_i$ and $S^g_i$ are the source terms.

The three-point Gauss-Hermite quadrature is employed in present work to get the discrete particle velocities in one dimension. The discrete velocities and associated weights in two dimension can be achieved by the tensor product method,
\begin{equation}
\bm{\xi}=\sqrt{3RT}
\Bigg{[}
\begin{aligned}
0&&1&&1&&0&&-1&&-1&&-1&&0&&1
\\
0&&0&&1&&1&&1&&0&&-1&&-1&&-1
\end{aligned}
\Bigg{]},
\
\omega_i =
	\begin{cases}
	\frac{4}{9},&i = 0\\
	\frac{1}{9},&i = 1,3,5,7\\
	\frac{1}{36},&i = 2,4,6,8
	\end{cases}.
\end{equation}
The equilibrium distribution function for $f_i^{eq}$ is expressed as
\begin{equation}
f_i^{eq} = \omega_i\phi(1+\bm{\xi_i}\cdot\bm{u}/c_s^2),
\label{Eq:phiEq}
\end{equation}
where $c_s = \sqrt{RT}$ is the sound speed. Ren\cite{Ren2016Improved} pointed out that by discarding high-order terms of velocity, the exact form of Allen-Cahn equation achieved through Chapman¨CEnskog analysis can be guaranteed. The source term $S_i^f$ consist of two parts and is defined as
\begin{equation}
S_i^f = \omega_i\theta\bm{\xi_i}\cdot\bm{n} + \omega_i\bm{\xi_i}\partial_t(\phi\bm{u})/RT.
\label{phiSource}
\end{equation}
The second part is necessary to eliminate the term of $\partial_t(\phi\bm{u})$ introduced via the C-E expansion\cite{Wang2016Comparative}.

The equilibrium distribution function for $g_i^{eq}$ is\cite{Zu2013Phase,Liang2014Phase}
\begin{equation}
g_i^{eq} =
	\begin{cases}
	\frac{p}{RT}(\omega_i-1)+\rho{\big(\Gamma_i(\bm{u})-\Gamma_i(0)\big)}, &i = 0,\\
	\frac{p}{RT}\omega_i+\rho{\big(\Gamma_i(\bm{u})-\Gamma_i(0)\big)}, &i\neq 0,
	\end{cases}
	\label{Eq:HydroEq}
\end{equation}
where
\begin{equation}
\Gamma_i(\bm{u})=\omega_i\Big[1 +\frac{\bm{\xi}_i\cdot\bm{u}}{RT} + \frac{(\bm{\xi}_i\cdot\bm{u})^2}{2(RT)^2}-\frac{\bm{u}\cdot\bm{u}}{2RT}\Big].
\end{equation}
The source term $S_i^{g}$ is defined as
\begin{equation}
S_i^{g} = \frac{(\bm{\xi}_i-\bm{u})}{RT}\cdot\left\{[\Gamma_i(\bm{u})-\Gamma_i(0)]\nabla(\rho{RT})+(\bm{F}_s+\bm{G})\Gamma_i(\bm{u})\right\}.
\label{HydroSource}
\end{equation}
It needs to be mentioned that Liang and Shi\cite{Liang2017Phase} proposed a simplified force model through discarding the term of $O(\delta_tMa^2)$ during the C-E analysis, which works well when the magnitude of flow velocity is relatively small. To keep rigorous and general property of the present scheme, however, the force model previously used by Liang\cite{Liang2014Phase} is applied. By choosing the appropriate expression for equilibrium distribution functions and source terms, we can get the exact macroscopic equations from the discrete kinetic equations via the C-E analysis, the details of which are shown in Appendix.

Since Eq.(\ref{fdke}) and Eq.(\ref{gdke}) share the same pattern, a new symbol $\psi$ is introduced to substitute either $f$ or $g$ for the convenience of illustration. Thus, the unified form of discrete kinetic equation is
\begin{equation}
\frac{\partial{\psi_i}}{\partial{t}}+\xi_i\cdot\nabla{\psi_i}=-\frac{\psi_i-\psi_i^{eq}}{\tau_f}+S^\psi_i.
\label{DKE}
\end{equation}
The DUGKS is applied to solve the above equation for its various advantages\cite{Wang2015A,Zhu2016Performance}. Integrating it on an control volume $V_j$ centered at $\bm{x}_j$ from time $t_n$ to $t_{n+1}$, we get
\begin{equation}
{\psi}_i^{n+1}-{\psi}_i^{n}+\frac{\Delta{t}}{\vert{V_j}\vert}J^{{\psi},n+1/2}=\frac{\Delta{t}}{2}[\Omega_i^{{\psi},n+1}+\Omega_i^{{\psi},n}]+\frac{\Delta{t}}{2}[S_i^{{\psi},n+1}+S_i^{{\psi},n}],
\label{DUGKSfundamental}
\end{equation}
where $\Omega_i^\psi = -(\psi_i-\psi_i^{eq})/{\tau_\psi}$, $V_j$ is the volume of cell with index $j$, $n$ is the time. $J^{\psi,n+1/2}$ is the microflux across the cell interface at the middle of current time interval with the following form
\begin{equation}
J^{\psi,n+1/2}=\int_{\partial{V_j}}(\bm{\xi}_i\cdot\bm{n})\psi_i(\bm{x}_f,\bm{\xi}_i,t_{n+1/2})d\bm{S},
\label{fluxE}
\end{equation}
where $\partial{V_j}$ is the surface of cell $V_j$, $\bm{n}$ is the outward unit vector normal to the surface element $dS$ and $\bm{x}_f$ denotes the position of surface element.
Trapezoidal rule is employed for the integration of the collision term $\Omega_i^{\psi}$ and source term $S_i^{\psi}$, and mid-point rule is chosen for the evaluation of the microflux $J^{\psi,n+1/2}$. To overcome the implicit treatment of source terms in Eq.(\ref{DUGKSfundamental}), two auxiliary distribution functions are introduced\cite{Guo2013Discrete,Guo2015Discrete}
\begin{subequations}
	\begin{equation}
	\tilde\psi_{i} = \psi_{i}-\frac{\Delta{t}}{2}\Omega_i-\frac{\Delta{t}}{2}S_i
	=\frac{2\tau_\psi+\Delta{t}}{2\tau_\psi}\psi_{i}-\frac{\Delta{t}}{2\tau_\psi}\psi_{i}^{eq}-\frac{\Delta{t}}{2}S_i,
	\label{Eq:Tilde}
	\end{equation}
	\begin{equation}
	\tilde\psi_{i}^{+} = \psi_{i}+\frac{\Delta{t}}{2}\Omega_i+\frac{\Delta{t}}{2}S_i
	=\frac{2\tau_\psi-\Delta{t}}{2\tau_\psi}\psi_{i}+\frac{\Delta{t}}{2\tau_\psi}\psi_{i}^{eq}+\frac{\Delta{t}}{2}S_i.
	\end{equation}
	\label{auxiliary}
\end{subequations}
Substitute Eq.(\ref{auxiliary}) into Eq.(\ref{DUGKSfundamental}) and rearrange each of these terms according to the time step, we have
\begin{equation}
\tilde{\psi}^{n+1}_i = \tilde{\psi}^{+,n}_i - \frac{\Delta{t}}{\vert{V_j}\vert}J^{\psi,n+1/2}.
\label{TrueDF}
\end{equation}
Instead of the original distribution function $\psi$, the auxiliary distribution function $\tilde{\psi}$ is updated. The key step to obtain an accurate $\tilde{\psi}$ lies in the evaluation of flux $J^{\psi,n+1/2}$. To get the original DF at intermediate moment of an time interval, Eq.(\ref{DUGKSfundamental}) is integrated  along its characteristic line within a half time step $h = \Delta{t}/2$,
\begin{equation}
\begin{split}
\psi_i(\bm{x}_f,t_n+h)-\psi_i(\bm{x}_f-\bm{\xi}_ih,t_n)&=\frac{h}{2}[\Omega_i^{\psi}(\bm{x}_f,t_n+h)+\Omega_i^{\psi}(\bm{x}_f-\bm{\xi}_ih,t_n)]\\
&+\frac{h}{2}[S_i^{\psi}(\bm{x}_f,t_n+h)+S_i^{\psi}(\bm{x}_f-\bm{\xi}_ih,t_n)].
\end{split}
\label{FluxOriginal}
\end{equation}
Again, to remove the implicit treatment of the collision term and source term, two auxiliary distribution functions are introduced
\begin{subequations}
\begin{equation}
\bar{\psi}_i = \psi_i-\frac{h}{2}\Omega_i^{\psi}-\frac{h}{2}S_i^{\psi}=\frac{2\tau_\psi+h}{2\tau_\psi}\psi_i-\frac{h}{2\tau_\psi}\psi_i^{eq}-\frac{h}{2}S_i^{\psi},
\end{equation}
\begin{equation}
\bar{\psi}_i^{+} = \psi_i+\frac{h}{2}\Omega_i^{\psi}+\frac{h}{2}S_i^{\psi}=\frac{2\tau_\psi-h}{2\tau_\psi}\psi_i+\frac{h}{2\tau_\psi}\psi_i^{eq}+\frac{h}{2}S_i^{\psi}.
\end{equation}
\end{subequations}
As a result, Eq.(\ref{FluxOriginal}) turns into
\begin{equation}
\bar{\psi}_i(\bm{x}_f,t_n+h)=\bar{\psi}_i^{+}(\bm{x}_f-\bm{\xi}_ih,t_n),
\label{ingenious}
\end{equation}
which is the most ingenious step in the evaluation of flux. An indispensable particle distribution function $\bar{\psi}_i(\bm{x}_f,t_n+h)$ that is related with the next half time step is reconstructed by tracing this set of particle back to the current time step, with the collision and force effect taken into consideration.

Two approaches, central scheme and upwind scheme, were put forward by Guo et al\cite{Guo2013Discrete,Guo2015Discrete} successively to obtain the value of $\bar{\psi}_i^{+}(\bm{x}_f-\bm{\xi}_ih,t_n)$. Since the simulation of two phase flow demands a low numerical dissipation on the algorithm, the central scheme is used in present study. After the update of auxiliary distribution function $\bar{\psi}_i(\bm{x}_f,t_n+h)$ and macroscopic variables located at the cell interface, the original distribution function can be calculated by
\begin{equation}
\psi_i=\frac{2\tau_\psi}{2\tau_\psi+h}\bar{\psi}_i+\frac{h}{2\tau_\psi+h}\psi_i^{eq} + \frac{\tau_\psi{h}}{2\tau_\psi+h}S_i^{\psi}.
\end{equation}
Thus, the flux $J^{\psi,n+1/2}$ can be evaluated from Eq.(\ref{fluxE}). And finally $\tilde{\psi}_i^{n+1}$ can be obtained according to Eq.(\ref{TrueDF}) with the two following equations
\begin{flalign}
\bar{\psi}_i^{+} &= \frac{2\tau_\psi-h}{2\tau_\psi+\Delta{t}}\tilde{\psi}_i+\frac{3h}{2\tau_\psi+\Delta{t}}\psi^{eq}_i+\frac{3\tau_\psi{h}}{2\tau_\psi+\Delta{t}}S_i^{\psi},\\
\tilde{\psi}_i^{+} &= \frac{4}{3}\bar\psi^{+}_i-\frac{1}{3}\tilde{\psi}_i.
\end{flalign}

The macroscopic variables including order parameter $\phi$,dynamic pressure $p$, and velocity $\bm{u}$ at each cell center are updated by
\begin{flalign}
\phi &= \sum_i{\tilde{f}_i}, \\
\bm{u} &= (\sum_i{\bm{\xi}_i\tilde{g}_i} + \frac{\Delta{t}}{2}\bm{F})/\rho, \\
p &= \frac{RT}{1-\omega_0}\Big[\sum_{i\neq{0}}\tilde{g}_{i}+\frac{\Delta{t}}{2}\bm{u}\cdot\nabla\rho+\rho[\Gamma_0(\bm{u})-\Gamma_0(0)]\Big].
\label{Eq:dynamicPressure}
\end{flalign}
The relaxation times are determined by the mobility and kinetic viscosity according to
\begin{equation}
\tau_f = M_\phi/RT,\ \tau_g = \mu/\rho{RT}.
\end{equation}
Generally two popular approaches are used in the calculation of the dynamic viscosity $\mu$. One approach is the linear interpolation of the reciprocals of the viscosities proposed by Zu and He\cite{Zu2013Phase}, i.e.,
\begin{equation}
\frac{1}{\mu}=\phi(\frac{1}{\mu_H}-\frac{1}{\mu_L})+\frac{1}{\mu_L}.
\end{equation}
The other common approach is to use a plain linear interpolation expressed as
\begin{equation}
{\mu}=\phi({\mu_H}-{\mu_L})+{\mu_L}.
\end{equation}
The reciprocal interpolation scheme shows a better accuracy while the plain interpolation scheme is able to enhance modeling stability. A detailed comparison between these two approaches is presented in Sec. \ref{SEC:BINARYFLOWTESTS}.\ref{SUBSEC:LayeredPoiseuilleflow}.

Since all of the tests in this paper use uniform Cartesian grid, a six-point numerical scheme\cite{Stiles2016High} is applied for the computation of $\nabla\phi$, in which the only gradient term needs to be updated. The Laplace operator is calculated by a general nine-point finite difference scheme to ensure the isotropic property. The temporal derivative in Eq.(\ref{phiSource}) is calculated by the first-order forward Euler scheme. The time step in present work is determined by the CFL condition, as follows:
\begin{equation}
\Delta{t}=CFL\frac{\delta_l}{\sqrt{3RT}}.
\end{equation}

It is worth noting that Zhang and Guo\cite{Zhang2018} proposed a quasi-incompressible model based on DUGKS. Here we give some remarks on the difference between the model of Guo and the present model. Firstly, the governing equations used to capture the interface in Guo's model is the Cahn-Hilliard (C-H) equation\cite{Cahn1958,Cahn1959} while the Allen-Cahn (A-C) equation\cite{Sun2007Sharp,Chiu2011,Geier2015Conservative} is employed in present model. Wang and Shi\cite{Wang2016Comparative} gives a detailed comparison between these two equations based on lattice Boltzmann (LB) model and the results show that the LB model for A-C equation gives more accurate and stable results. Secondly, the mass equations used to describe the flow field are different. In Guo's model, a quasi-incompressible model\cite{Yang2016Lattice} is adopted, with which the uniform mass conservation can be guaranteed while the divergence-free velocity is introduced in present model and mass conservation can only be ensured in the single-phase zone. The phenomenon of mass generation or consumption can be observed in the mixing layer when a density contrast exists, as is explained in the first part of this section. Lastly, the number of first-order derivative terms and second-order derivative terms that need to be updated during the iterative process is different. For Guo's model, three first-order derivative terms including $\nabla{\rho},\nabla{p},\nabla{\mu_\phi}$ and two second-order derivative terms covering $\Delta{\mu_\phi}$ and $\Delta{\phi}$ needs to be calculated during each iterative process due to the introduction of the quasi-incompressible model. Since $\mu_\phi$ itself contains a second order derivative term, a fourth order derivative term needs to be calculated to obtain an accurate $\Delta{\mu_\phi}$, actually. Those miscellaneous but indispensable derivatives may be responsible for the relatively low density ratio (no larger than 10) in each of their tests. In terms of the present method, only two derivative terms, $\nabla{\phi}$ and $\Delta{\phi}$, are necessary during the process of calculation. The maximum density ratio can reach as much as 1000 in the stationary bubble case. In brief, Guo's method gives a more accurate description about mass transfer during the process of phase transition at the price of introducing more spatial derivative terms up to the fourth order and is incapable of dealing with high density ratio scenarios. The present method offers a more efficient and concise way in interface capturing and behaves well at a relatively high density ratio case except that mass conservation cannot be guaranteed in the mixing zone of a two phase flow, which is a common problem within the framework of the phase-field theory under the assumption of incompressibility condition.

\section{INTERFACE-CAPTURING TESTS}\label{SEC:INTERFACE-CAPTURING TESTS}
In this section, four typical benchmark problems, including interface diagonal translation, Zalesak's disk rotation, interface elongation and interface deformation, are used to validate the interface-capturing ability of the present scheme. Each of the velocity fields is specified in advance, hence only Eq. (\ref{Eq:interfaceEq}) needs to be solved. The dimensionless parameters, P\'{e}clet number and Cahn number, are defined as\cite{Ren2016Improved}
\begin{equation}
Pe=\frac{U_0L_0}{M_\phi},Cn = \frac{W}{L_0},
\end{equation}
where $U_0$ is the reference velocity and $L_0$ is the side length of computational domain. The grid size $\delta_l$ is kept at unity and the CFL number remains at 0.5 unless otherwise specified. To quantitatively evaluate the performance of the present method and make a comparison with the results of LB method, the $L_2$-norm based error of the order parameter is used\cite{Liang2014Phase}:
\begin{equation}
E_\phi=\frac{\sum_{\bm{x}}\vert\phi(\bm{x},T)-\phi(\bm{x},0)\vert^2}{\sum_{\bm{x}}\vert\phi(\bm{x},0)\vert^2}.
\end{equation}
\subsection{Interface diagonal translation}
A circular interface with radius $R = L_0/4$ is settled at the center of a square domain with $L_0$$\times$${L_0}$ cells. Periodic boundary condition is applied to all of its sides. The uniform velocity field is specified as
\begin{equation}
u(x,y) = U_0,v(x,y)=U_0.
\end{equation}
The circular interface would move back to its initial location after $T=L_0/U_0$ time. A comparison based on the convergence rate between the current scheme and the LB method\cite{Liang2017Phase} is provided. The effects of P\'{e}clet number and velocity are also investigated.

\begin{figure}[H]
\centering
\includegraphics[width=0.48 \textwidth]{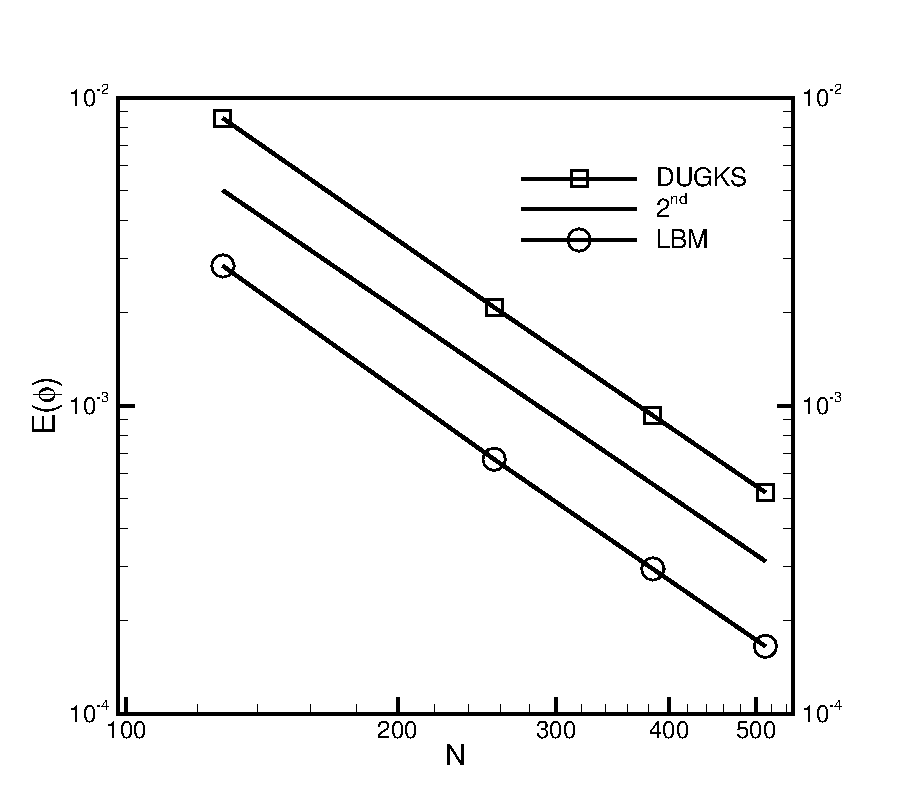}
\caption{Convergence rate of DUGKS and LBM, $Pe = 128$, $Cn = 1/32$, $M_\phi = 0.02$, $\Delta{t}= 0.5$}
\label{figureL2ofPhi}
\end{figure}

To obtain the convergence rate, the grid number along each side of the square is refined from 128 to 512. In order to keep P\'{e}clete number, Cahn number and mobility coefficient constant, the reference velocity $U_0$ and interface width $W$ are tuned along with the variation of grid number. Comparison results shown in Fig. \ref{figureL2ofPhi} exhibit a second-order convergence accuracy of both DUGKS and LB method. The overall error yielded by DUGKS is a bit higher than that of LB method, for DUGKS has a relatively larger numerical dissipation.\cite{Wang2016Comparison}.

The reference velocity is used to tune the P\'{e}clet number to investigate its effect on the relative error. The other parameters, including the mobility coefficient, reference length and Cahn number, are kept at a constant value. When it comes to the effect of mobility coefficient, also the reference velocity is tuned to keep the P\'{e}clet number fixed. The reference length keeps a constant value of 256 and the Cahn number is $4/256$. Results pertaining to $Pe$ and $M_\phi$ are presented in Table \ref{TAB:diagTransPe} and \ref{TAB:diagTransMPhi}, respectively.
\begin{table}[H]
   \caption{$L_2$ Error of $\phi$ for interface diagonal translation, $M_\phi = 0.02, L_0 = 256, Cn = 4/256$}
   \begin{ruledtabular}
   \begin{tabular}{ccccc}
    \toprule
    $Pe$ & 128 & 256 & 512 & 1024 \\
    \midrule[0.4pt]
    DUGKS & 3.6470E-3 & 5.7916E-3 & 1.1548E-2 & 2.3409E-2 \\
    LBM & 1.9808E-3 & 1.9769E-3 & 1.9669E-3 & 1.9173E-3 \\
    \bottomrule
   \end{tabular}
   \end{ruledtabular}
   \label{TAB:diagTransPe}
\end{table}
\begin{table}[H]
   \caption{$L_2$ Error of $\phi$ for interface diagonal translation, $Pe = 256, L_0 = 256,Cn = 4/256$}
   \begin{ruledtabular}
   \begin{tabular}{ccccc}
    \toprule
    $M_\phi$ & 0.02 & 0.04 & 0.064 & 0.1\\
    \midrule[0.4pt]
    DUGKS & 5.7916E-3 & 5.1422E-3 & 5.3437E-3 & 6.4816E-3 \\
    LBM & 1.9769E-3 & 1.9720E-3 & 2.0981E-3 & 2.8430E-3 \\
    \bottomrule
   \end{tabular}
   \end{ruledtabular}
   \label{TAB:diagTransMPhi}
  \end{table}

It can be shown in Table \ref{TAB:diagTransPe} that the relative error of DUGKS continues increasing as the P\'{e}clet number goes up gradually while the results of LBM always remain at the same level. The differences is caused by the reconstruction method adopted in the evaluation of flux. As is mentioned in section \ref{sec:subsecDUGSKforTwoPhase}, central scheme instead of upwind scheme is used to ensure the low numerical dissipation. Since the P\'{e}clet number indicates the ratio of the rate of advection by the rate of mobility driven by an gradient, the flow is mainly dominated by advection when the P\'{e}clet number is relatively large. Surely the central scheme used in flux evaluation would cause deviations under such a circumstance. To overcome this problem, a high-order upwind scheme needs to be developed. Table \ref{TAB:diagTransMPhi} gives the relative error of the order parameter for DUGKS and LBM at $Pe = 256$. Both of these two methods give stable results with an magnitude order as low as $1.0 \times 10^{-3}$ despite the increase of mobility.

In the scenario of uniform velocity field, the performance of DUGKS fails to compare with that of LB method at a relatively large P\'{e}clet number, while DUGKS is able to give results comparable to LB method when the P\'{e}clet number is relatively low.
\subsection{Zalesak's disk rotation}
\begin{figure}[H]
\centering
\includegraphics[width=0.48 \textwidth]{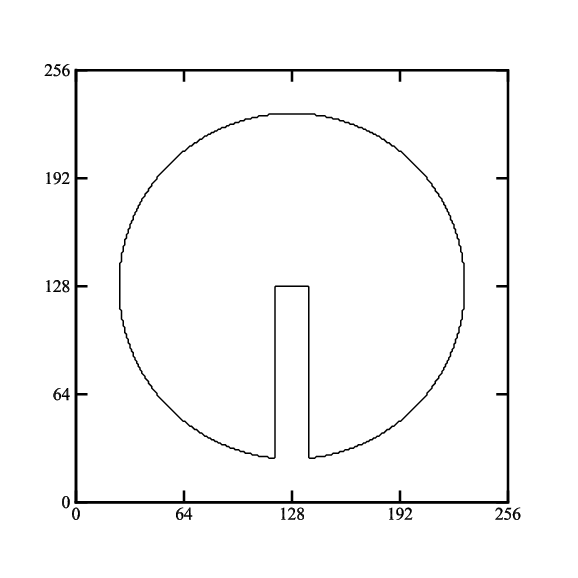}
\caption{Initial state of Zalesak's disk, $Cn = 4/256$, $M_\phi = 0.02$}
\label{FIG:ZalesakT0}
\end{figure}

\begin{figure}[H]
\centering
\subfigure[DUGKS]
{
\includegraphics[width=0.48 \columnwidth]{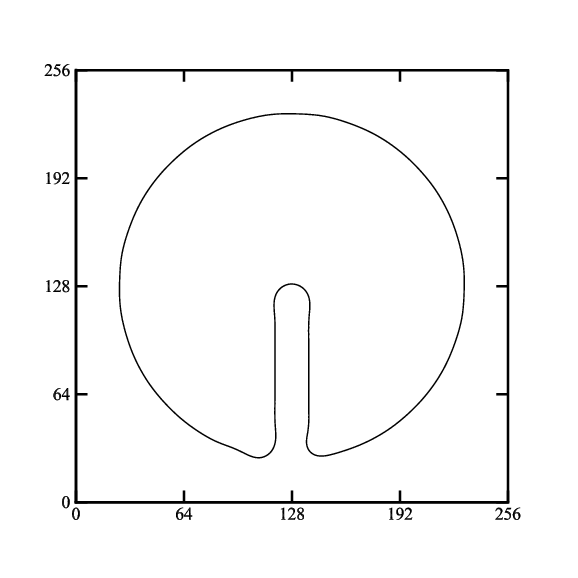}
\label{FIG:ZalesakDUGKSPe128}
}
\subfigure[LBM]
{
\includegraphics[width=0.48 \columnwidth]{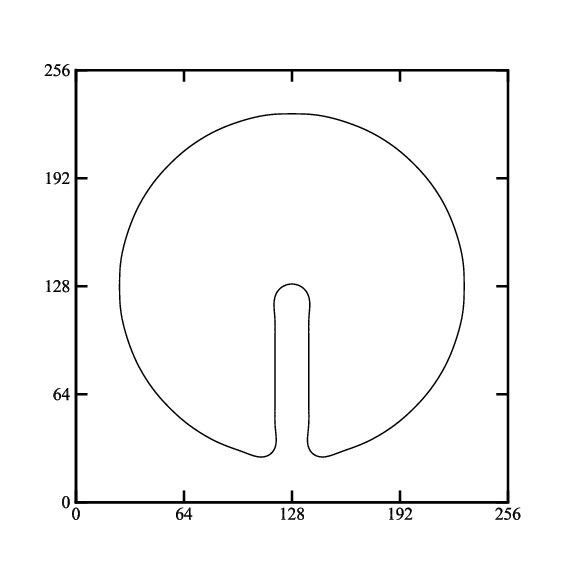}
\label{FIG:ZalesakLBMPe128}
}
\caption{Results of Zalesak's disk after one period at $Pe = 128$}
\label{FIG:ZalesakPe128}
\end{figure}

\begin{figure}[H]
\centering
\subfigure[DUGKS]
{
\includegraphics[width=0.48 \columnwidth]{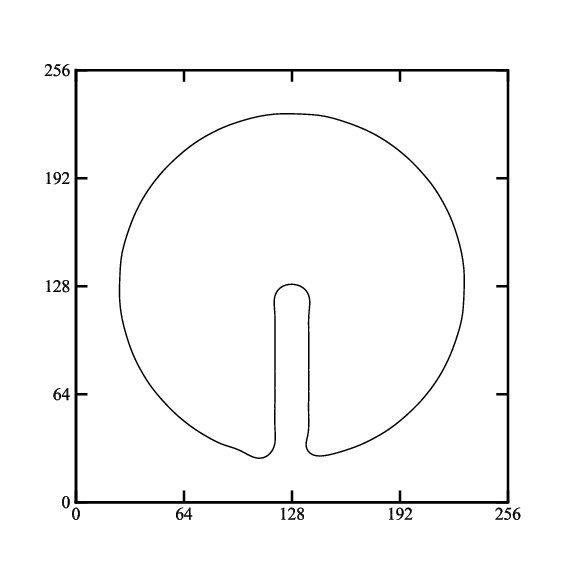}
\label{FIG:ZalesakDUGKSPe256}
}
\subfigure[LBM]
{
\includegraphics[width=0.48 \columnwidth]{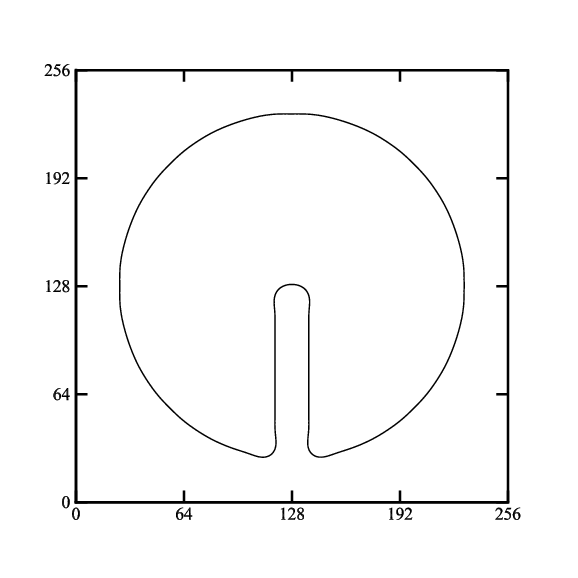}
\label{FIG:ZalesakLBMPe256}
}
\caption{Results of Zalesak's disk after one period at $Pe = 256$}
\label{FIG:ZalesakPe256}
\end{figure}

\begin{figure}[H]
\centering
\subfigure[DUGKS]
{
\includegraphics[width=0.48 \columnwidth]{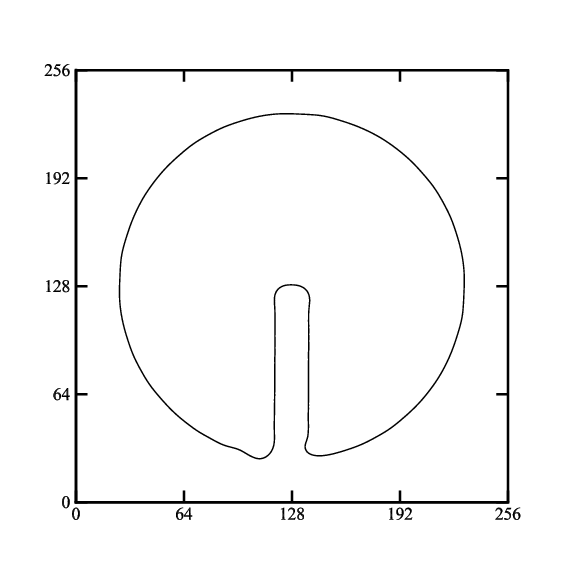}
\label{FIG:ZalesakDUGKSPe512}
}
\subfigure[LBM]
{
\includegraphics[width=0.48 \columnwidth]{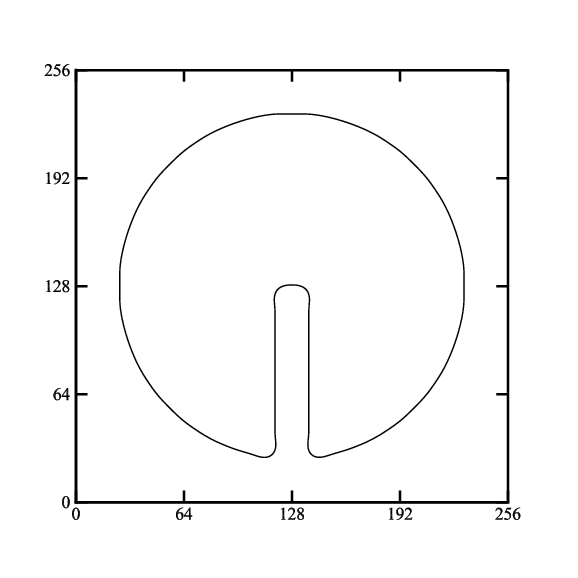}
\label{FIG:ZalesakLBMPe512}
}
\caption{Results of Zalesak's disk after one period at $Pe = 512$}
\label{FIG:ZalesakPe512}
\end{figure}

\begin{figure}[H]
\centering
\subfigure[DUGKS]
{
\includegraphics[width=0.48 \columnwidth]{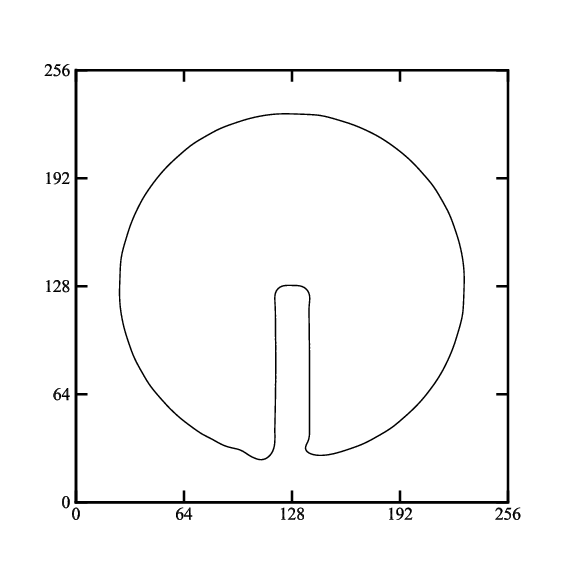}
\label{FIG:ZalesakDUGKSPe1024}
}
\subfigure[LBM]
{
\includegraphics[width=0.48 \columnwidth]{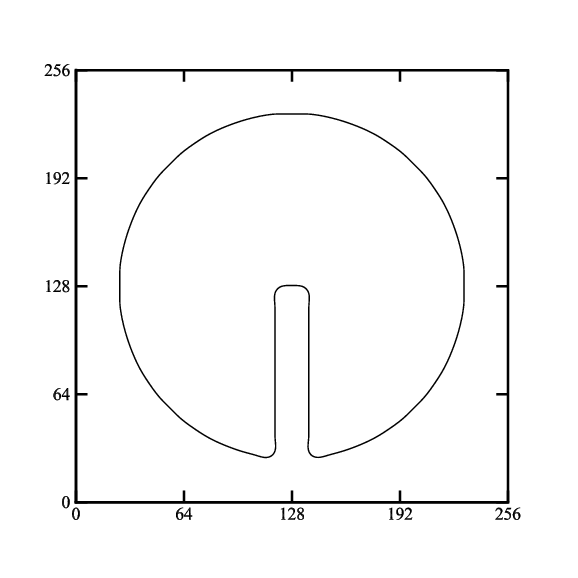}
\label{FIG:ZalesakLBMPe1024}
}
\caption{Results of Zalesak's disk after one period at $Pe = 1024$}
\label{FIG:ZalesakPe1024}
\end{figure}

As shown above, the diagonal translation of circular interface does not involve any sharp interface. To further test the ability of present method in capturing sharp interface, Zalesak's disk\cite{Zalesak1979,Ding2007,Zu2013Phase} test is conducted. The disk with a slot is initially located at the center of a $256$$\times$${256}$ square domain, as illustrated in Fig. \ref{FIG:ZalesakT0}. The radius of the disk is set as 100 and the width of the slot is 20. The disk is driven by a irrotational flow field governed by
\begin{equation}
\begin{aligned}
u(x,y) &= -\frac{U_0{\pi}}{L_0}(y-0.5L_0),\\
v(x,y) &= \frac{U_0{\pi}}{L_0}(x-0.5L_0).
\end{aligned}
\end{equation}
In theory, the disk would return to its initial position after $T = 2L_0/U_0$ time. The Cahn number is fixed at $4/256$. The order parameter inside the disk is initialized by $\phi_H$ and $\phi_L$ accounts for the rest part. As a diffusive interface method, a transition layer is necessary for the description of the interface. However, no smooth function is available at start time to generate this transition layer in current case. Thus, there exists discontinuities in the vicinity of the step-shaped interface.

we first make a comparison of the interface patterns obtained respectively with DUGKS and LBM after one period time, which are shown in Fig. \ref{FIG:ZalesakPe128}-\ref{FIG:ZalesakPe1024}. Both DUGKS and LBM can give a stable evolution of the interface and no sawteeth phenomenon\cite{Liang2014Phase} is observed. At low P\'{e}clet numbers (128 and 256), DUGKS is able to capture the interface as accurate as the LBM. As the p\'{e}clet number increases, a small stretch of the sharp corners at the tip of the slot can be observed, which is mainly caused by the flux scheme chosen in current DUGKS.

To give a quantitative analysis on the results of DUGKS and LBM, the relative error of the order parameter in terms of P\'{e}clet number is presented in Table {\ref{TAB:ZalesakPe}}. It can be found that the results achieved by both methods in the end show a large deviation from the initial distribution of $\phi$. As mentioned above, no smooth function is adopted to define the interface. Hence the interface width is zero at initial time. However, a interface with nonzero width is formed during the process of evolution. That is why a relatively large deviation between the results achieved at initial and final moment exists. This viewpoint can also explain the same phenomenon shown in Table \ref{TAB:ZalesakMPhi}, in which results concerning the effect of mobility coefficient is presented.
\begin{table}[H]
   \caption{$L_2$ Error of $\phi$ for Zalesak's disk rotation, $M_\phi = 0.02, L_0 = 256, Cn = 4/256$}
   \begin{ruledtabular}
   \begin{tabular}{cccccc}
    \toprule
    $Pe$ & 128 & 256 & 512 & 640 & 1024\\
    \midrule[0.4pt]
    DUGKS & 1.085E-1 & 1.071E-1 & 1.078E-1 & 1.089E-1 & 1.118E-1\\
    LBM & 1.069E-1 & 1.057E-1 & 1.048E-1 & 1.044E-1 &1.041E-1\\
    \bottomrule
   \end{tabular}
   \end{ruledtabular}
   \label{TAB:ZalesakPe}
\end{table}
\begin{table}[H]
   \caption{$L_2$ Error of $\phi$ for Zalesak's disk rotation, $Pe = 256, L_0 = 256,Cn = 4/256$}
   \begin{ruledtabular}
   \begin{tabular}{cccccc}
    \toprule
    $M_\phi$ & 0.02 & 0.04 & 0.064 & 0.08 & 0.1\\
    \midrule[0.4pt]
    DUGKS & 1.071E-1 & 1.068E-1 & 1.065E-1 & 1.068E-1 & 1.070E-1\\
    LBM & 1.057E-1 & 1.048E-1 & 1.058E-1 & 1.059E-1 &1.061E-1\\
    \bottomrule
   \end{tabular}
   \end{ruledtabular}
   \label{TAB:ZalesakMPhi}
  \end{table}
\subsection{Interface elongation in a shear flow}\label{Subsec:elongationinterface}
\begin{figure}[H]
\centering{}
\subfigure[DUGKS]
{
\includegraphics[width=0.48 \columnwidth]{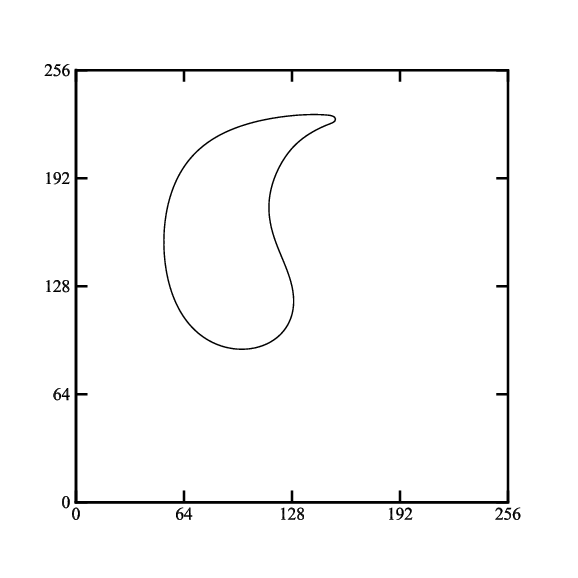}
\label{FIG:ShearStretchDUGKS0_25T}
}
\subfigure[LBM]
{
\includegraphics[width=0.48 \columnwidth]{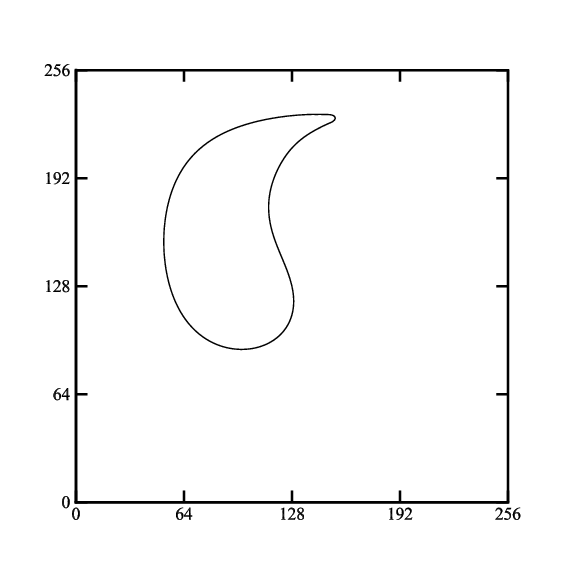}
\label{FIG:ShearStretchLBM0_25T}
}
\caption{Results of interface elongation in a shear flow at $t = 0.25T$, $M_\phi = 0.01$, $Pe = 256$}
\label{FIG:ShearStretch0_25T}
\end{figure}

\begin{figure}[H]
\centering
\subfigure[DUGKS]
{
\includegraphics[width=0.48 \columnwidth]{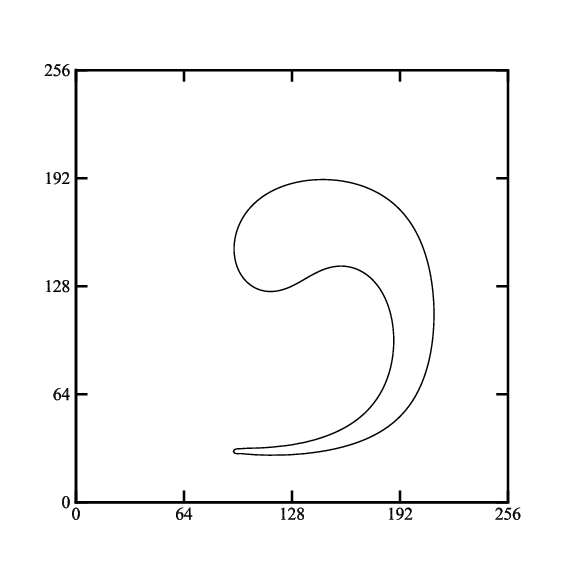}
\label{FIG:ShearStretchDUGKS0_5T}
}
\subfigure[LBM]
{
\includegraphics[width=0.48 \columnwidth]{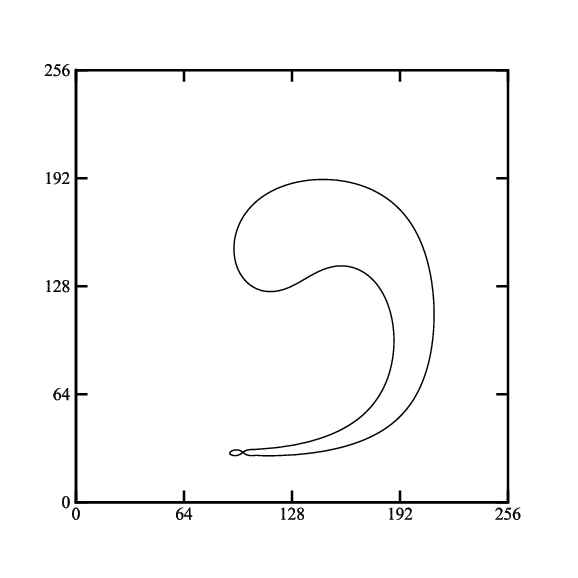}
\label{FIG:ShearStretchLBM0_5T}
}
\caption{Results of interface elongation in a shear flow at $t = 0.5T$, $M_\phi = 0.01$, $Pe = 256$}
\label{FIG:ShearStretch0_55T}
\end{figure}

\begin{figure}[H]
\centering
\subfigure[DUGKS]
{
\includegraphics[width=0.48 \columnwidth]{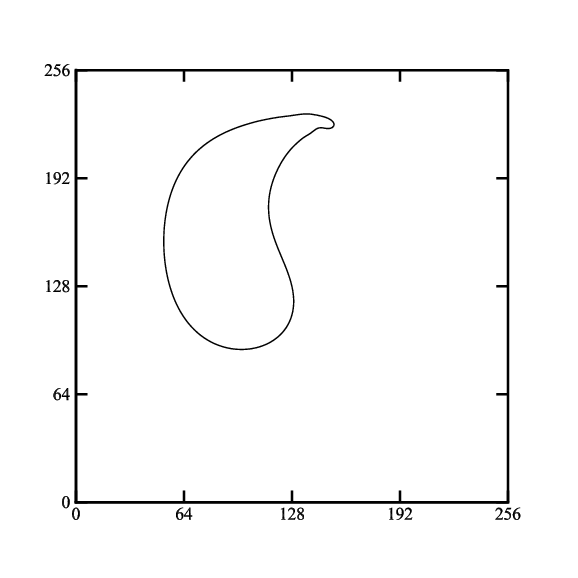}
\label{FIG:ShearStretchDUGKS0_75T}
}
\subfigure[LBM]
{
\includegraphics[width=0.48 \columnwidth]{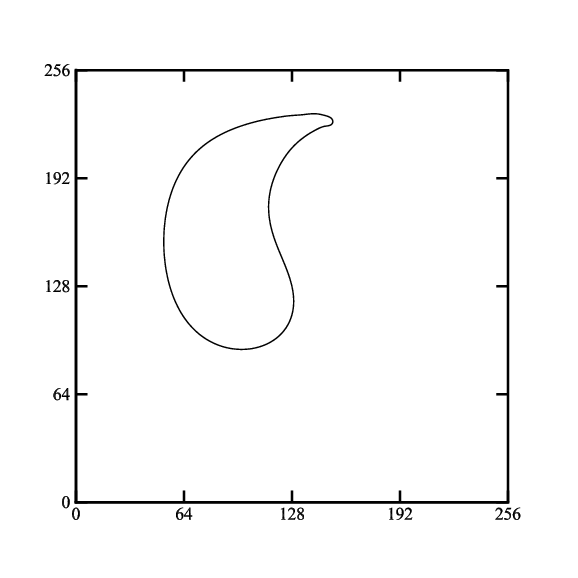}
\label{FIG:ShearStretchLBM0_75T}
}
\caption{Results of interface elongation in a shear flow at $t = 0.75T$, $M_\phi = 0.01$, $Pe = 256$}
\label{FIG:ShearStretch0_75T}
\end{figure}

\begin{figure}[H]
\centering
\subfigure[DUGKS]
{
\includegraphics[width=0.48 \columnwidth]{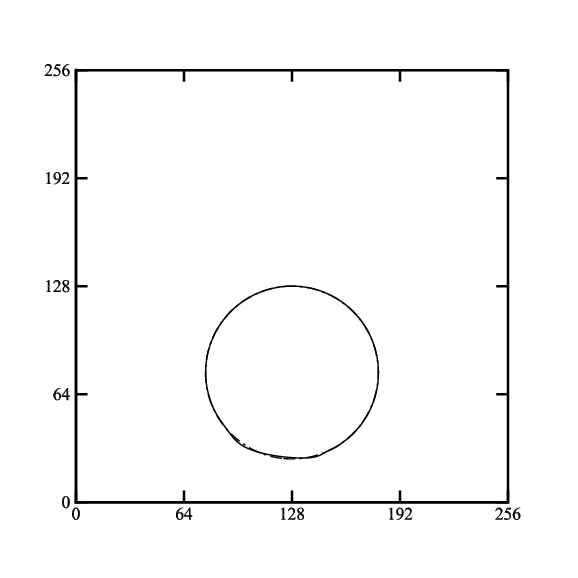}
\label{FIG:ShearStretchDUGKS1T}
}
\subfigure[LBM]
{
\includegraphics[width=0.48 \columnwidth]{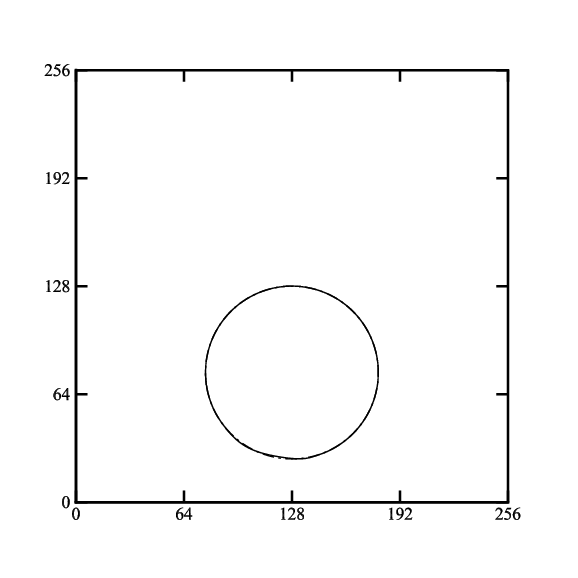}
\label{FIG:ShearStretchLBM1T}
}
\caption{Results of interface elongation in a shear flow at $t = T$, $M_\phi = 0.01$, $Pe = 256$}
\label{FIG:ShearStretch1T}
\end{figure}

\begin{figure}[H]
\centering
\subfigure[DUGKS]
{
\includegraphics[width=0.48 \columnwidth]{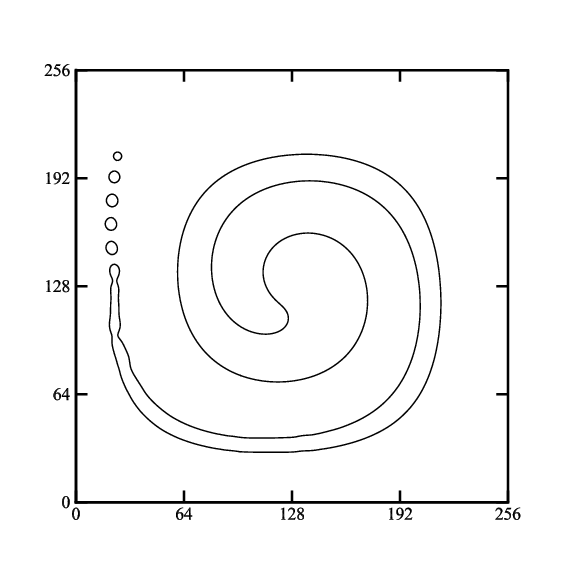}
\label{FIG:ShearStretchDUGKS0_5TLiang2014}
}
\subfigure[LBM]
{
\includegraphics[width=0.48 \columnwidth]{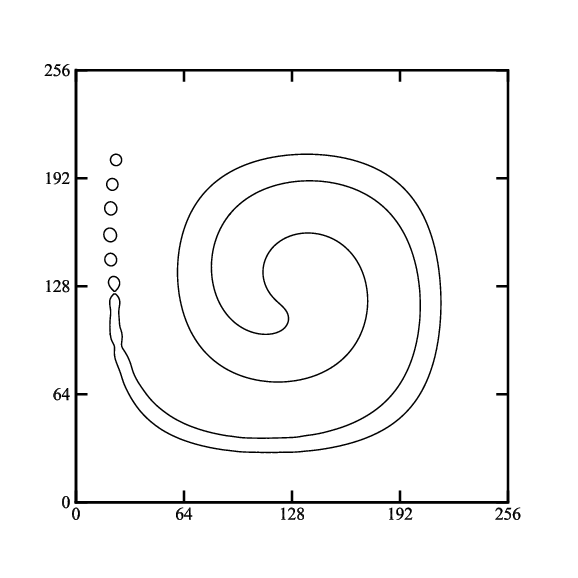}
\label{FIG:ShearStretchLBM0_5TLiang2014}
}
\caption{Results of interface elongation in a shear flow at $t = 0.5T$, $M_\phi = 0.1$, $Pe = 256$}
\label{FIG:ShearStretchLiang2014}
\end{figure}

Neither of the above two cases deal with large shape deformation as the interface remains unchanged during the evolution process. To further validate the ability of present method in capturing interface deformation, the circular interface elongation in a shear flow is studied. In this case, a circular interface with radius $R = L_0/5$ is initially placed at $x=0.5L_0$ and $y = 0.3L_0$ in a square domain with $L_0$$\times$${L_0}$ cells, where $L_0=256$ is the reference length. The velocity field is governed by
\begin{equation}
\begin{aligned}
u(x,y) &= U_0{\pi}sin(\frac{\pi{x}}{L_0})cos(\frac{\pi{y}}{L_0}),\\
v(x,y) &= -U_0{\pi}cos(\frac{\pi{x}}{L_0})sin(\frac{\pi{y}}{L_0}).
\end{aligned}
\end{equation}
After $L_0/U_0$ time, the velocity field is reversed to its opposite direction. In this way, the elongated interface would recover to its initial state after another $L_0/U_0$ time. The whole time used in this process is defined as the time period, $T=2L_0/U_0$. The interface is displayed by the contour level of $\phi=0.5(\phi_H+\phi_L)$. Fig. \ref{FIG:ShearStretch0_25T}-\ref{FIG:ShearStretch1T} illustrate the stretching process of the interface obtained by the present method and LBM. At the time of $t = 0.5T$, the tail tip of the stretched interface is about to break in LBM while DUGKS is able to maintain this tail tip stable. The velocity field is reversed afterwards and the stretched interface starts to recover. At $t = 0.75T$, the contour of interface shown in Fig. \ref{FIG:ShearStretch0_75T} is approximate to the results presented in Fig. \ref{FIG:ShearStretch0_25T} except a small distortion at the tip of stretched interface. After a period time, the stretched interface is restored back up to its original pattern, which is shown in Fig. \ref{FIG:ShearStretch1T}. A close inspection towards the results in Fig. \ref{FIG:ShearStretch1T} shows that there exists a slight deviation at the lower-left part of the interface between the final (solid line) and initial (dash dotted line) moment, which originates from the tip distortion during the process of restoration. To give an quantitative description about this deviation, the relative error of order parameter in terms of the P\'{e}clet number is calculated and presented in Table \ref{TAB:ShearStretchPe}. As the P\'{e}clet number increases, a tiny increment can be observed in the relative error obtained with DUGKS while results achieved with LBM remain stable. It is worth noting that at a relatively large p\'{e}clet number DUGKS fails to give a result comparable to that of LBM in the case of interface diagonal translation. In current test, however, the results obtained by DUGKS and LBM at a large p\'{e}clet number are pretty close. The effect of mobility coefficient is also studied and results are presented in Table \ref{TAB:ShearStretchPhi}. A same growth trend can be observed in the relative errors obtained with both DUGKS and LBM as the mobility coefficient increases. Also results produced by DUGKS show good agreement with that of LBM at various mobility coefficients.
\begin{table}[H]
   \caption{$L_2$ Error of $\phi$ for interface stretch in a shear flow, $M_\phi = 0.01, L_0 = 256, Cn = 4/256$}
   \begin{ruledtabular}
   \begin{tabular}{cccccc}
    \toprule
    Pe &  256 & 512 & 1024 & 1638 \\
    \midrule[0.4pt]
    DUGKS & 2.596E-2 & 2.831E-2 & 3.246E-2 & 3.798E-2 \\
    LBM & 1.405E-2 & 1.232E-2 & 1.309E-2 & 1.653E-2\\
    \bottomrule
   \end{tabular}
   \end{ruledtabular}
   \label{TAB:ShearStretchPe}
\end{table}
\begin{table}[H]
   \caption{$L_2$ Error of $\phi$ for interface stretch in a shear flow, $Pe = 256, L_0 = 256,Cn = 4/256$}
   \begin{ruledtabular}
   \begin{tabular}{cccccc}
    \toprule
    $M_\phi$ & 0.01 & 0.02 & 0.04 & 0.05 & 0.064\\
    \midrule[0.4pt]
    DUGKS & 2.596E-2 & 2.681E-2 & 2.749E-2 & 2.868E-2 & 3.072E-2\\
    LBM & 1.405E-2 & 1.480E-2  & 1.561E-2 &  1.635E-2 & 1.796E-2\\
    \bottomrule
   \end{tabular}
   \end{ruledtabular}
   \label{TAB:ShearStretchPhi}
  \end{table}
A further study on the capacity of the kinetic model based on Allen-Cahn equation is conducted with a time dependent velocity field, which is governed by the same equations used by Liang\cite{Liang2014Phase}. With such a velocity field, the interface will stretch out to a greater length. Results at the moment of half period are shown in Fig. \ref{FIG:ShearStretchLiang2014}. It can be seen clearly that the tail tip of the interface breaks up into small drops. As has been illustrated by Liang\cite{Liang2014Phase}, the kinetic model based on Cahn-Hilliard equation is able to capture the long tails and the initial circular shape of the interface can be recovered accurately. The deficiency in the Allen-Cahn-based kinetic model may have serious impacts on numerical simulations referring to subtle interface changes.

\subsection{Interface deformation in a smoothed shear flow}
\begin{figure}[H]
\centering{}
\subfigure[DUGKS]
{
\includegraphics[width=0.48 \columnwidth]{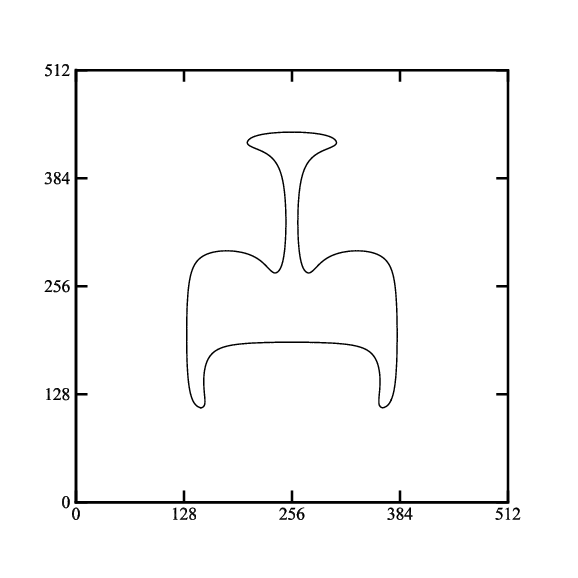}
\label{FIG:DeformSmoothDUGKS0_25T}
}
\subfigure[LBM]
{
\includegraphics[width=0.48 \columnwidth]{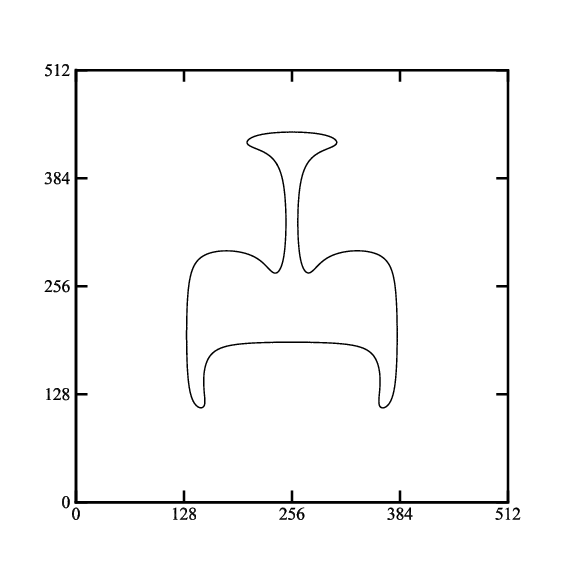}
\label{FIG:DeformSmoothLBM0_25T}
}
\caption{Results of interface deformation in a smoothed flow at $t = 0.25T$, $M_\phi = 0.02$, $Pe = 2048$}
\label{FIG:DeformSmooth0_25T}
\end{figure}

\begin{figure}[H]
\centering{}
\subfigure[DUGKS]
{
\includegraphics[width=0.48 \columnwidth]{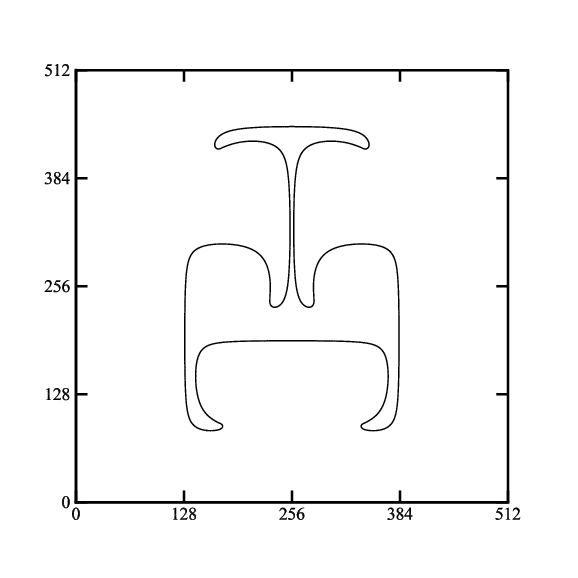}
\label{FIG:DeformSmoothDUGKS0_5T}
}
\subfigure[LBM]
{
\includegraphics[width=0.48 \columnwidth]{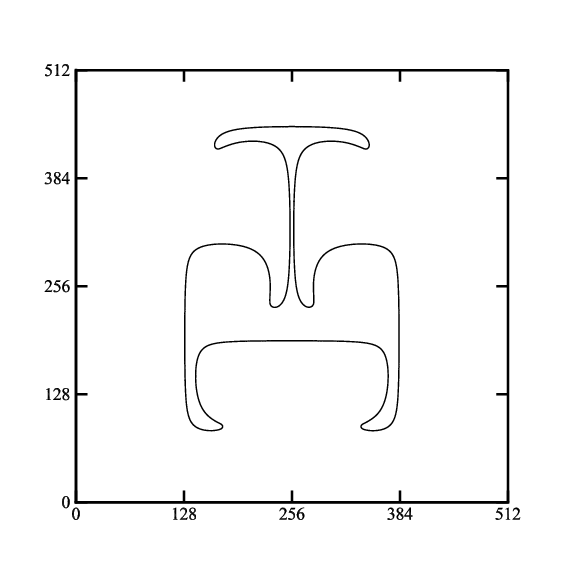}
\label{FIG:DeformSmoothLBM0_5T}
}
\caption{Results of interface deformation in a smoothed flow at $t = 0.5T$, $M_\phi = 0.02$, $Pe = 2048$}
\label{FIG:DeformSmooth0_5T}
\end{figure}

\begin{figure}[H]
\centering{}
\subfigure[DUGKS]
{
\includegraphics[width=0.48 \columnwidth]{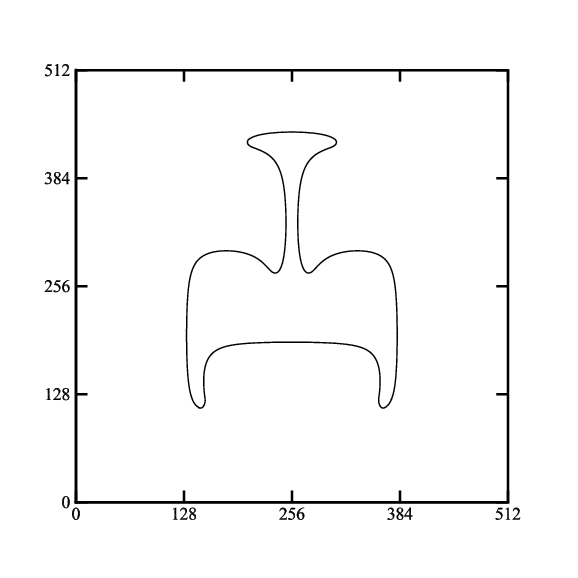}
\label{FIG:DeformSmoothDUGKS0_75T}
}
\subfigure[LBM]
{
\includegraphics[width=0.48 \columnwidth]{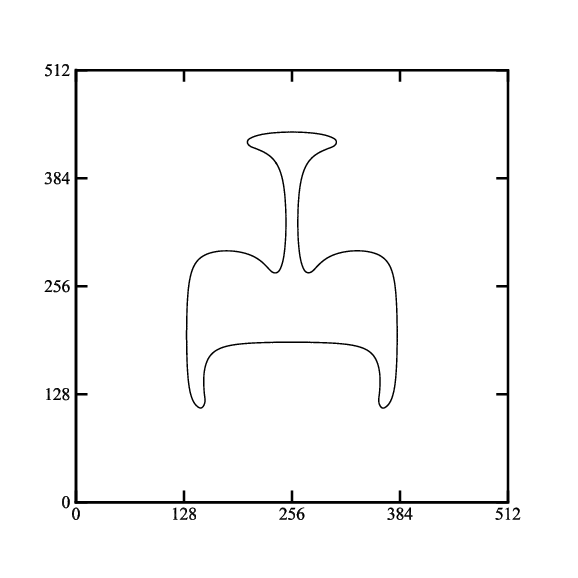}
\label{FIG:DeformSmoothLBM0_75T}
}
\caption{Results of interface deformation in a smoothed flow at $t = 0.75T$, $M_\phi = 0.02$, $Pe = 2048$}
\label{FIG:DeformSmooth0_75T}
\end{figure}

\begin{figure}[H]
\centering{}
\subfigure[DUGKS]
{
\includegraphics[width=0.48 \columnwidth]{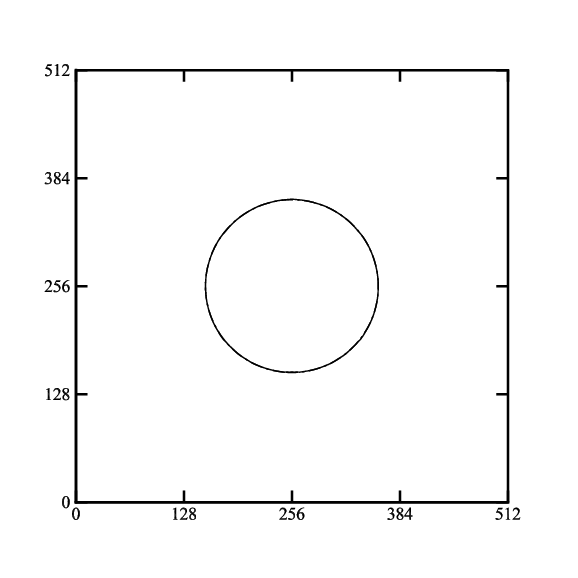}
\label{FIG:DeformSmoothDUGKS1T}
}
\subfigure[LBM]
{
\includegraphics[width=0.48 \columnwidth]{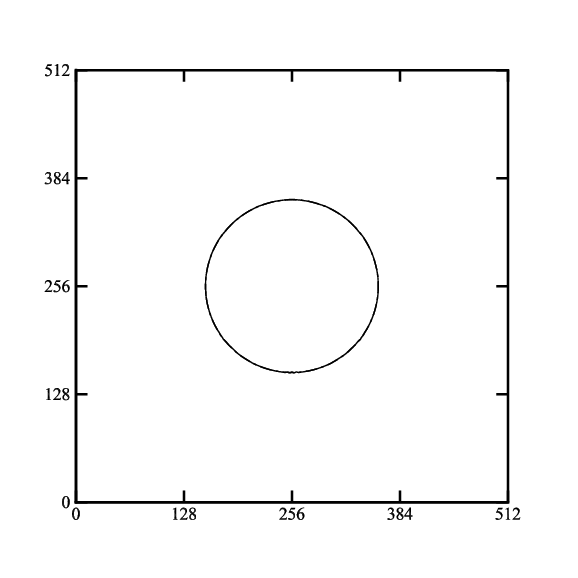}
\label{FIG:DeformSmoothLBM1T}
}
\caption{Results of interface deformation in a smoothed flow at $t = T$, $M_\phi = 0.02$, $Pe = 2048$}
\label{FIG:DeformSmooth1T}
\end{figure}

To further explore the ability of present method in capturing interface deformation, we continue to conduct another test about the deformation of circular interface in a smoothed shear flow, which is regarded as one of the most rigorous problems as the interface undergoes a severe deformation\cite{Liang2014Phase,Geier2015Conservative}. The circular interface with a radius $R = L_0/5$ is located at the center of a periodic domain with $L_0$$\times$$L_0$ cells, where $L_0 = 512$ is the reference length. The velocity field is controlled by
\begin{equation}
\begin{aligned}
u(x,y) &= -U_0sin(\frac{4{\pi}x}{L_0})sin(\frac{4{\pi}y}{L_0})cos(\frac{{\pi}t}{T}),\\
v(x,y) &= -U_0cos(\frac{4{\pi}x}{L_0})cos(\frac{4{\pi}y}{L_0})cos(\frac{{\pi}t}{T}),
\end{aligned}
\end{equation}
where $U_0$ is the reference velocity and $T = L_0/U_0$ is the period. As in the above case, the interface undergoes transfiguration in the first half period and reconsolidation in the last half period. The main difference is that a temporal smoothing term $cos(\frac{{\pi}t}{T})$ is introduced in current case to avoid the rapid shift of velocity field. The deformation of the interface driven by a smooth velocity field is shown in Fig. \ref{FIG:DeformSmooth0_25T}-\ref{FIG:DeformSmooth1T}. It can be found that the results obtained with DUGKS share the same deformation pattern as those of LBM. The restored interface(solid line) after one period time overlaps with the initial one(dash dotted line) exactly for both methods. In addition, quantitative comparisons between DUGKS and LBM are presented in Table \ref{TAB:DeformationPe} and \ref{TAB:DeformationPhi}, illustrating the effects of the P\'{e}clet number and mobility coefficient, respectively. It can be shown clearly that even at a large P\'{e}clet number, the results provided by DUGKS are almost the same as that of LBM, which is mainly attributed to the finer mesh resolution. As the mobility coefficient increases, DUGKS even shows a better performance than that of LB method.
\begin{table}[H]
   \caption{$L_2$ Error of $\phi$ for interface deformation in a smoothed flow, $M_\phi = 0.02, L_0 = 512, Cn = 4/512$}
   \begin{ruledtabular}
   \begin{tabular}{cccccc}
    \toprule
    $Pe$ & 512 & 1024 & 1638 &2048\\
    \midrule[0.4pt]
    DUGKS & 9.762E-3 & 7.986E-3 & 1.033E-2 & 1.339E-2 \\
    LBM & 9.409E-3 & 7.281E-3 & 9.026E-3 & 1.203E-2\\
    \bottomrule
   \end{tabular}
   \end{ruledtabular}
   \label{TAB:DeformationPe}
\end{table}
\begin{table}[H]
   \caption{$L_2$ Error of $\phi$ for interface deformation in a smoothed flow, $Pe = 512, L_0 = 512,Cn = 4/512$}
   \begin{ruledtabular}
   \begin{tabular}{cccccc}
    \toprule
    $M_\phi$ & 0.01 & 0.02 & 0.04 & 0.064 & 0.08\\
    \midrule[0.4pt]
    DUGKS & 1.051E-2 & 9.762E-3 & 8.652E-3 & 7.880E-3 & 7.570E-3\\
    LBM &  9.422E-3 &  9.409E-3  & 9.355E-3 &  9.255E-3 & 9.178E-3\\
    \bottomrule
   \end{tabular}
   \end{ruledtabular}
   \label{TAB:DeformationPhi}
  \end{table}
\section{BINARY FLOW TESTS}\label{SEC:BINARYFLOWTESTS}
As the ability of DUGKS in interface capturing has been validated, more tests related to hydrodynamic behavior are used to explore the capacity of present method. In this section, four benchmark cases including stationary bubble, layered Poiseuille flow, spinodal decomposition and Rayleigh-Taylor instability are chosen to test and verify the performance of current model. The convergence criterion for steady flows depends on
\begin{equation}
\frac{\sum_{\bm{x}}\vert{Q(\bm{x},n+1000)}-{Q(\bm{x},n)}\vert^2}{\sum_{\bm{x}}\vert{Q(\bm{x},n+1000)}\vert^2} < 1.0 \times 10^{-8},
\label{Eq:Error}
\end{equation}
where $Q$ stands for either the order parameter $\phi$ or the flow velocity $\bm{u}$ and $n$ is the time step. The CFL number remains at 0.5 if not otherwise specified.

\subsection{Stationary bubble}\label{stationarybubble}
The stationary bubble is a basic problem in verifying the newly developed numerical method\cite{Fakhari2010Phase,LI2019}. At initial state, a light bubble immersed in the heavy liquid is placed at the center of a square domain with $L_0$$\times$$L_0$ cells. Periodic boundary conditions are applied to all boundaries. The initial profile of order parameter is give by
\begin{equation}
\phi=\frac{\phi_H+\phi_L}{2} + \frac{\phi_H-\phi_L}{2}\times\text{tanh}\frac{2[\sqrt{(x-x_c)^2+(y-y_c)^2}-R]}{W},
\end{equation}
where $(x_c,y_c)$ is the center of the computational domain and $R$ is the bubble radius. The interface width $W$ is fixed at 5 and the kinetic viscosity $\nu$ remains at 0.1 for the whole flow field. The density ratio varies from 10 to 1000 for different results. The end condition is determined by Eq.(\ref{Eq:Error}) with $Q$ replaced with $\phi$.

\begin{figure}[H]
\centering{}
\includegraphics[width=0.48 \columnwidth]{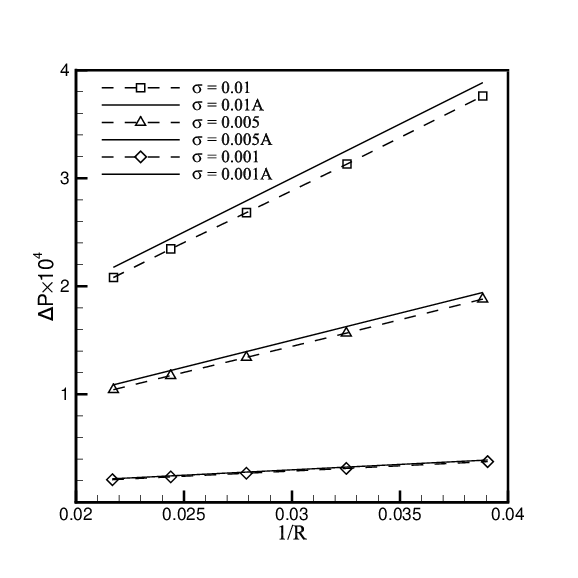}
\caption{Validation of Laplace's law with a density ratio of 1000 and $M_\phi = 0.1$}
\label{FIG:LaplaceLaw}
\end{figure}

\begin{figure}[H]
\centering{}
\includegraphics[width=0.48 \columnwidth]{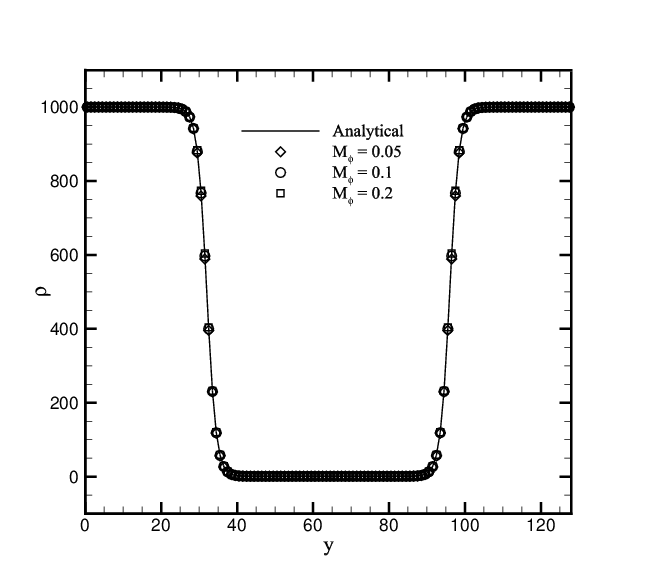}
\caption{Density profile along the vertical center line with a density ratio of 1000}
\label{FIG:RhoProfile}
\end{figure}

\begin{figure}[H]
\centering{}
\includegraphics[width=0.48 \columnwidth]{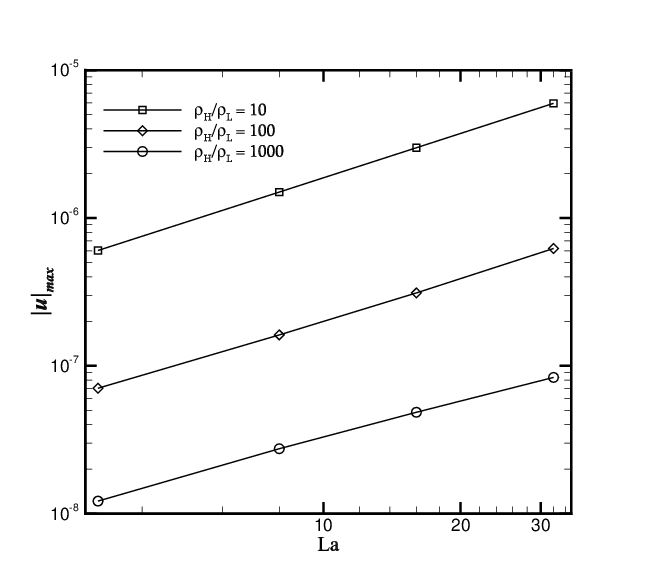}
\caption{Maximum magnitude of spurious velocity with various La number and density ratio}
\label{FIG:Umax-La}
\end{figure}

The performance of present method is firstly examined by Laplace's law. The relationship between pressure difference across the interface and reciprocal of bubble radius is determined by $\Delta{P} = \sigma/R$, where $P$ is the thermodynamic pressure and is calculated through $P = p_0-\kappa\phi\Delta\phi+\kappa\vert\nabla\phi\vert^2/2+p$ with the equation of state $p_0 = \phi\partial_\phi\epsilon(\phi)-\epsilon(\phi)$\cite{Liang2014Phase,Zheng2015Lattice,Yang2016Lattice,Zhang2018}. Fig. \ref{FIG:LaplaceLaw} presents the validation of Laplace's law based on current method at a density ratio of 1000. As the surface tension coefficient (STC) increases, obvious deviations between the numerical results (solid line) and analytical results (dash line with symbols) can be observed. For all situations, the ratio of numerical STC to analytical ones is around 96.5\%, which is approximated to the results of Liang\cite{Liang2014Phase} obtained with an MRT model. The absolute error between the numerical results and analytical results is enlarged with the growth of STC.
Fig. \ref{FIG:RhoProfile} depicts the density profile along the vertical center line with various values of mobility coefficient. It can be seen that numerical results match with the analytical solution exactly, which indicates the fundamental ability of current method in simulation of hydrodynamic problems.
The magnitude of spurious velocity draws attention to numbers of researchers focusing on two phase problems\cite{Yu2010Multirelaxation,Guo2011Force,Ba2016Multiple}. Here we give a detailed study on the relationship among density ratio, Laplace number (La) and spurious velocity. The Laplace number is defined as $\sigma\rho_L{R}/\mu_L^2$, which is different from Liang\cite{Liang2017Phase}. Fig. \ref{FIG:Umax-La} shows the maximum magnitude of spurious velocity at various La number and density ratio. It can be seen that the maximum magnitude of spurious velocity is linear with La number regardless of the density ratio. Actually the Laplace number is adjusted with the variation of STC in current test. Other parameters such as the radius, the density and dynamic viscosity of light phase are kept at constants. Hence we can conclude that a linear relationship between the maximum magnitude of spurious velocity and STC is observed. As the density ratio goes up, a significant drop in the maximum magnitude of spurious velocity can be found. The same trend can be observed in Liang's work\cite{Liang2017Phase}. The maximum magnitude of spurious velocity is no less than $10^{-6}$ in terms of other previous LB models\cite{Yu2010Multirelaxation,Liang2014Phase,Ba2016Multiple}. The above results testify that the current method is able to produce lower spurious velocities.

\subsection{Layered Poiseuille flow}\label{SUBSEC:LayeredPoiseuilleflow}
The layered Poiseuille flow is used as a benchmark in the validation of various two-phase approaches\cite{Wang2015Multiphase,Ren2016Improved,LIANG2017efficient}. Two immiscible fluids are driven by a constant body force $\bm{G} = (G_x,0)$ in a infinite channel. The upper region of $0 < y {\leqslant} h$ in the channel is filled with the fluid of small viscosity while the other part is filled with the fluid of large viscosity. Periodical boundary conditions are applied to the inlet and outlet and no-slip boundary condition is implemented on the upper and lower walls. When the flow reaches its steady state, the velocity field are in consistent with
\begin{equation}
u(y)=\left\{
\begin{aligned}
\frac{G_xh^2}{2\mu_L}\Big[-\Big(\frac{y}{h}\Big)^2-\frac{y}{h}\Big(\frac{\mu_L-\mu_H}{\mu_L+\mu_H}\Big)+\frac{2\mu_L}{\mu_L+\mu_H}\Big],&\ 0\ <\ y\ \ {\leqslant}\ h, \\
\frac{G_xh^2}{2\mu_H}\Big[-\Big(\frac{y}{h}\Big)^2-\frac{y}{h}\Big(\frac{\mu_L-\mu_H}{\mu_L+\mu_H}\Big)+\frac{2\mu_H}{\mu_L+\mu_H}\Big],&\ -h\ {\leqslant}\ y\ {\leqslant}\ 0.
\end{aligned}
\right.
\end{equation}

\begin{figure}[H]
\centering{}
\subfigure[$\mu^{*} = 10$]
{
\includegraphics[width=0.48 \columnwidth]{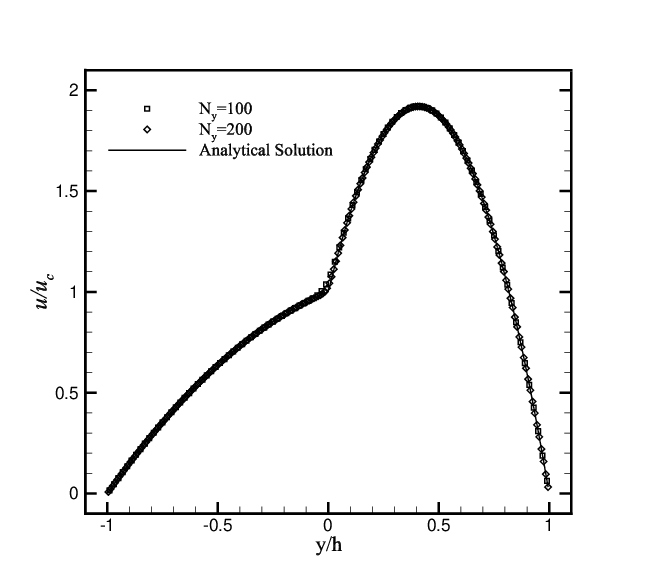}
\label{FIG:uMu10}
}
\subfigure[$\mu^{*} = 100$]
{
\includegraphics[width=0.48 \columnwidth]{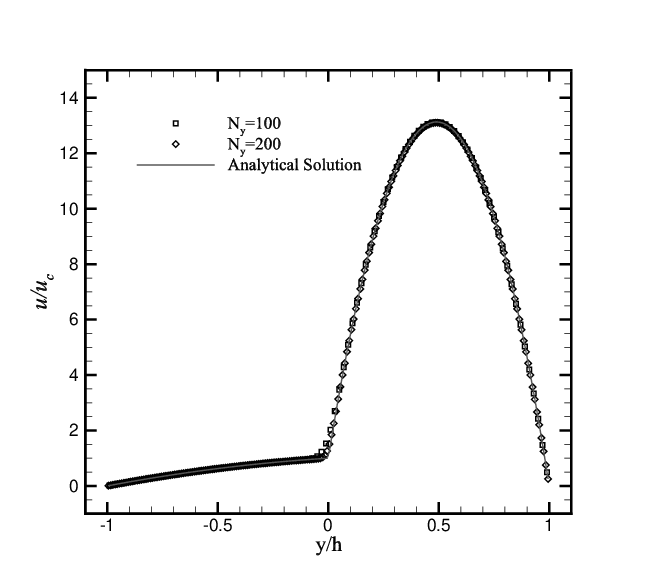}
\label{FIG:uMu100}
}
\subfigure[$\mu^{*} = 1000$]
{
\begin{minipage}[b]{1.0\textwidth}
\centering{}
\includegraphics[width=0.48 \columnwidth]{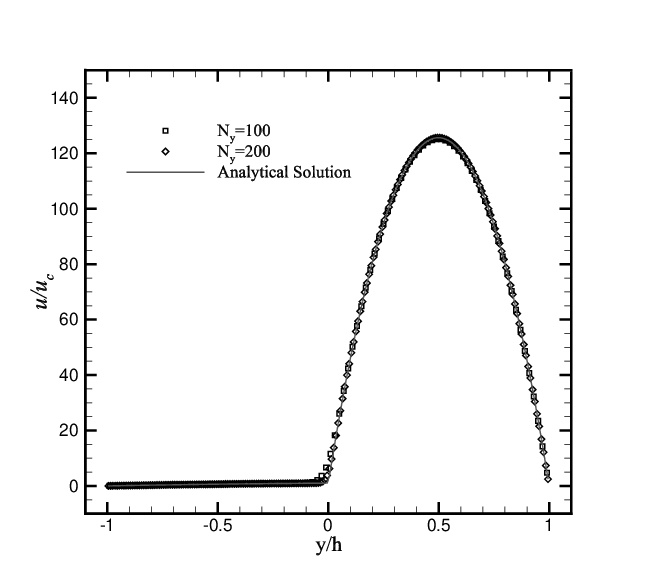}
\end{minipage}
\label{FIG:uMu1000}
}
\caption{Velocity profile of layered Poiseuille flow with various viscosity ratios}
\label{FIG:Poiseuille}
\end{figure}
\begin{figure}[H]
\centering{}
\includegraphics[width=0.5 \columnwidth]{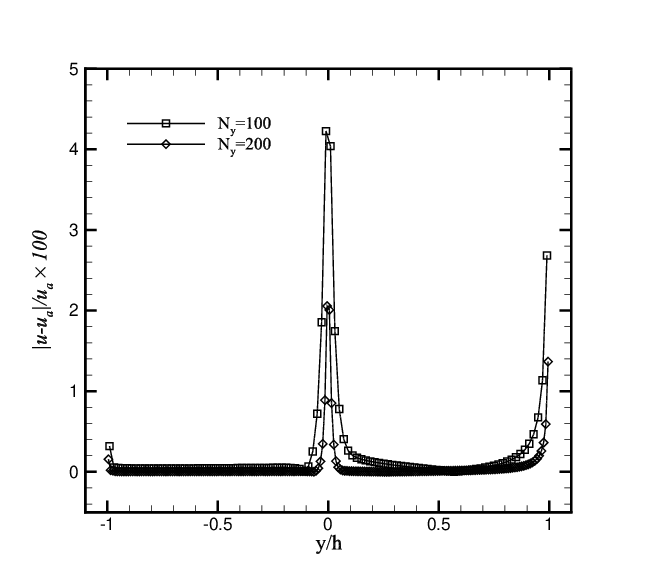}
\caption{Relative errors of layered Poiseuille flow with $\mu^* = 10$}
\label{FIG:uErrorMu10}
\end{figure}
\begin{figure}[H]
\centering{}
\subfigure[ ]
{
\includegraphics[width=0.48 \columnwidth]{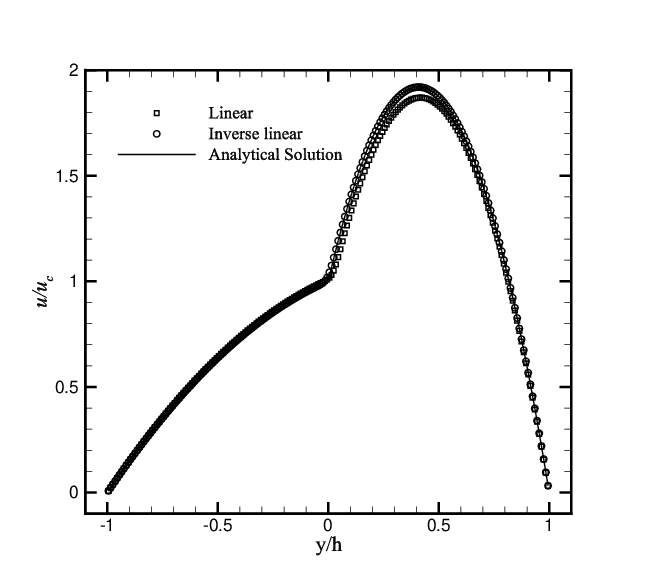}
\label{FIG:uMuMu}
}
\subfigure[ ]
{
\includegraphics[width=0.48 \columnwidth]{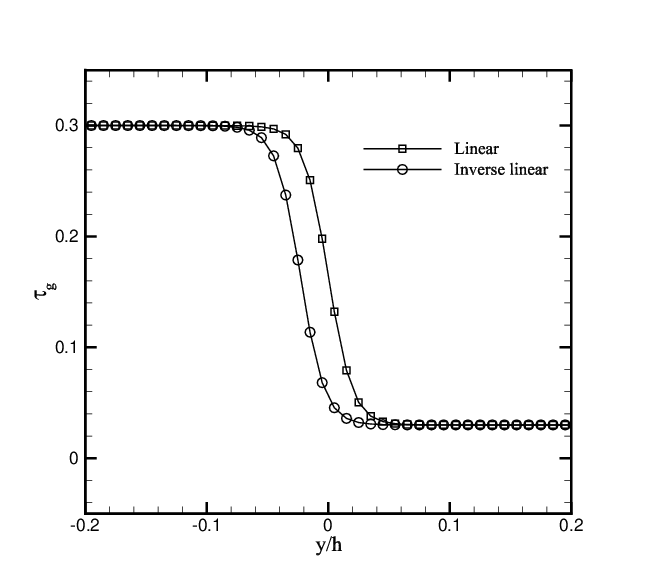}
\label{FIG:uTauTau}
}
\caption{Velocity profile (a) and relaxation time (b) with different interpolation schemes ($\mu^* = 10$)}
\label{FIG:MuMuTauTau}
\end{figure}

The central velocity $u_c$ at steady state is related to the constant driving force $G_x$, i.e., $u_c=G_xh^2/(\mu_L+\mu_H)$. The computing process terminates when Eq. (\ref{Eq:Error}) is satisfied, with $Q$ replaced by $u$. Two sets of grid, $10{\times}100$ and $10{\times}200$, are used in our simulation and comparisons are made between the numerical results and analytical ones. The central velocity $u_c$ is set to be $10^{-4}$, which is small enough to guarantee the incompressible condition. Other parameters are set as $W = 4,\sigma = 10^{-3},\ \rho_H = \rho_L = 1$. Three conditions with different dynamic viscosity ratio are considered in current case. The results of velocity profile are normalized by the central velocity. Fig. \ref{FIG:Poiseuille} presents the velocity profile along $y$ direction with various viscosity ratios ${\mu^*}$. It can be found that results obtained with both sets of grid are in good agreement with the analytical solution. The deviation mainly occurs at the interfacial region. As the mesh resolution goes finer, the deviation becomes smaller. A quantitatively description of the relative error between the numerical and analytical solution is presented in Fig. \ref{FIG:uErrorMu10}. Compared to the results of Ren\cite{Ren2016Improved}, the relative error achieved by DUGKS is a bit larger. This is mainly caused by the different schemes used in the evaluation of dynamic viscosity. To avoid the diffusion effect at the interfacial region, Ren adopts a step function for the dynamic viscosity while in our study, a continuous function is implemented. The overall $L_2$-norm error is recorded and it is found that the maximum value of the numerical error is $4.866E-03$ at the condition of $\mu^*=1000,N_y=200$ and $1.431E-02$ at the condition of $\mu^* = 1000, N_y = 100$, which have the same order as those in the literature\cite{Zheng2015Lattice}. The results above shows that present method is accurate enough in simulations involving large viscosity ratio.
In some of others work\cite{Fakhari2017Improved,Liang2017Phase}, a linear interpolation of the original dynamic viscosities is used to estimate the viscosity at the interfacial region. Comparisons between the profiles of velocity as well as relaxation time obtained through these two schemes are made and shown in Fig. \ref{FIG:MuMuTauTau}. It can be seen that linear scheme underestimate the velocity profile of large viscosity fluid. As is demonstrated by Zu and He\cite{Zu2013Phase}, the scheme of inverse linear interpolation ensures the continuity of viscosity flux at the mixture layer. It is reasonable to get a better result using this scheme. Liang and Shi\cite{LIANG2017efficient} also adopts the scheme of inverse linear interpolation for the estimation of interfacial viscosity in their newly research.

\subsection{Spinodal decomposition}

Spinodal decomposition, also known as phase separation, is a pervasive phenomenon in the study of immiscible fluids. It occurs due to the existence of fluctuations in a homogeneous mixture with a metastable state. Several studies on the spinodal decomposition problem have been carried out with the implementation of LB methods\cite{Xu2003Phase,Zu2013Phase,Gan2015Discrete,Liang2017Phase}. Among above works, only Liang\cite{Liang2017Phase} preforms a simulation with a large density ratio of 1000. To demonstrate the capability of current method in the study of phase separation process under high density ratio and to further illustrate the mass diffusion phenomenon during this process, the same spinodal decomposition problem is investigated. Our simulation is carried out in a square domain with a mesh of $200 \times 200$. Periodic boundary conditions are implemented at all boundaries. The initial distribution of the order parameter is defined by
\begin{equation}
\phi(x,y)=0.6 + \text{rand}(x,y),
\end{equation}
where $\text{rand}(x,y)$ is a random function used to impose fluctuations on the homogeneous mixture. The density field is calculated by Eq. (\ref{Eq:rho}), where $\rho_H$ and $\rho_L$ are set to be 1000 and 1, respectively. The kinetic viscosity ratio of $\nu_L$ to $\nu_H$ is fixed at 10. Other parameters are given as $W = 4, \sigma = 0.1$, and $M_\phi=0.1$. The dimensionless evolution time during the phase separation process is defined as $t^{*}=t/T$, where $T = \rho_H\nu_H{W}/\sigma$. The termination moment of our simulation is set at $t = 2500$, which is long enough to prove the stability of current method\cite{Zu2013Phase}. Fig. \ref{FIG:Evolution} depicts several contours of density distribution at various moments extracted from the process of phase separation. At a preliminary state, the small fluctuations in density evolve into large-scale inhomogeneities and interfaces separating different phases are beginning to emerge. Then the inhomogeneities drives the material of light phase into tiny bubbles with irregular shapes. As the system develops, some of these bubbles keep on coalescing into large ones. Eventually, a thermodynamic equilibrium state at which binary phases with distinctive contrast can be observed is reached.
\begin{figure}[H]
\centering{}
\subfigure[$t^{*} = 1.25$]
{
\includegraphics[width=0.48 \columnwidth]{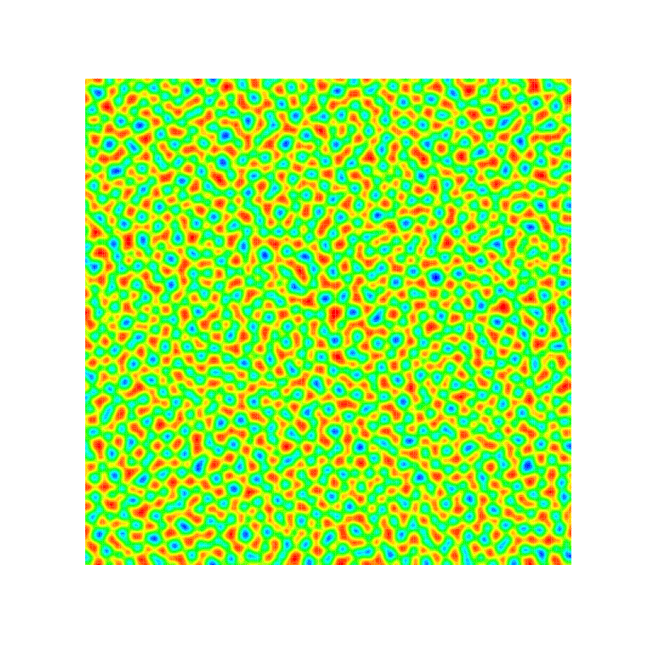}
\label{FIG:step1K}
}
\subfigure[$t^{*} = 5$]
{
\includegraphics[width=0.48 \columnwidth]{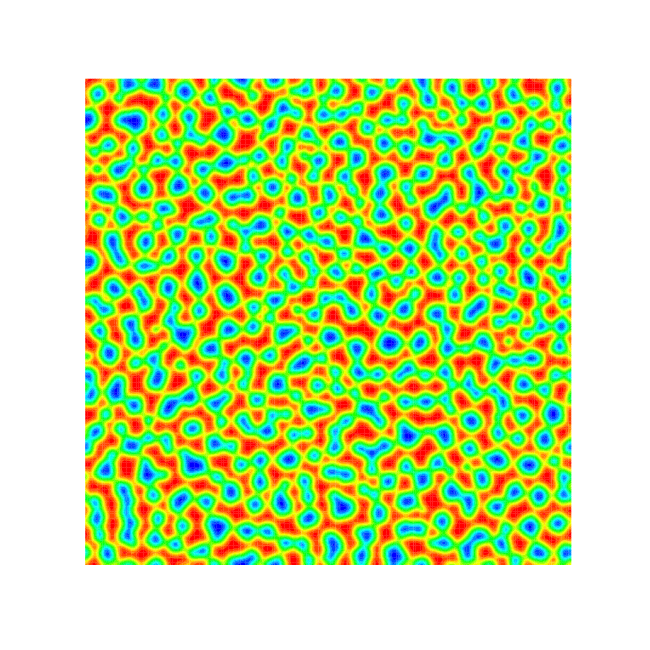}
\label{FIG:step4K}
}
\label{FIG:SpinodalD14K}
\end{figure}
\addtocounter{figure}{-1}
\begin{figure}[H]
\addtocounter{figure}{1}
\centering{}
\subfigure[$t^{*} = 6.25$]
{
\includegraphics[width=0.48 \columnwidth]{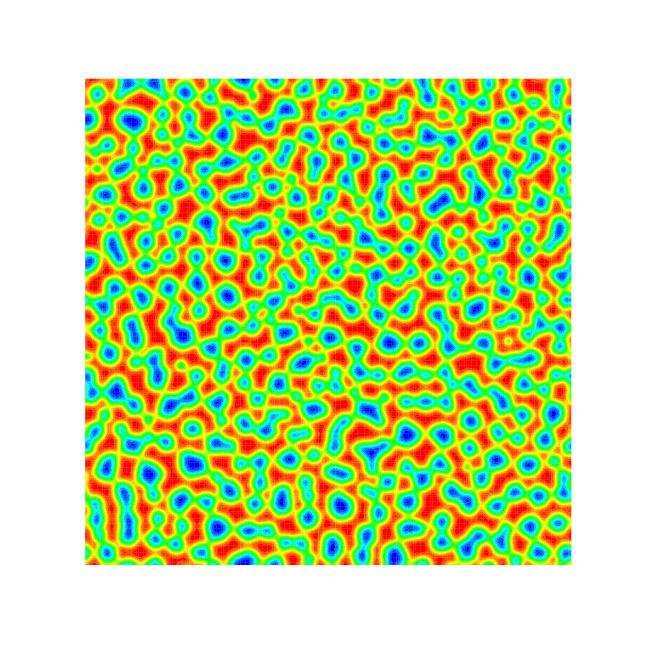}
\label{FIG:step5K}
}
\subfigure[$t^{*} = 62.5$]
{
\includegraphics[width=0.48 \columnwidth]{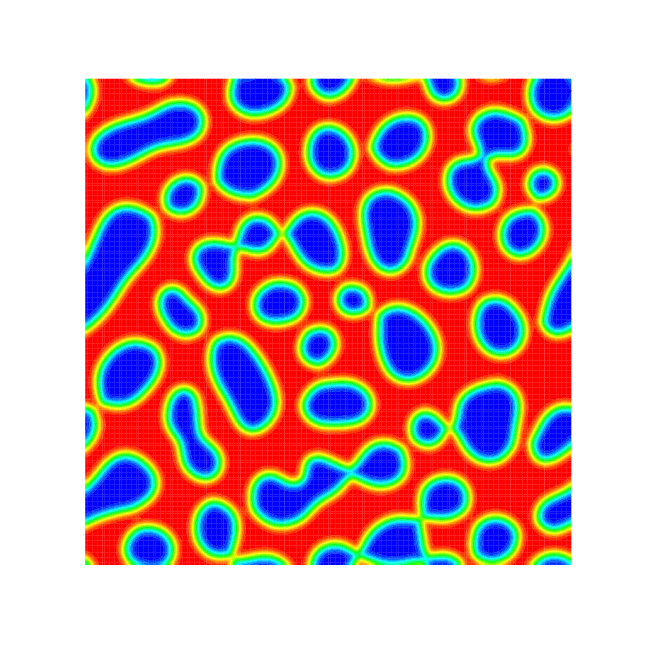}
\label{FIG:step50K}
}
\end{figure}
\addtocounter{figure}{-1}
\begin{figure}[H]
\addtocounter{figure}{1}
\centering{}
\subfigure[$t^{*} = 312.5$]
{
\includegraphics[width=0.48 \columnwidth]{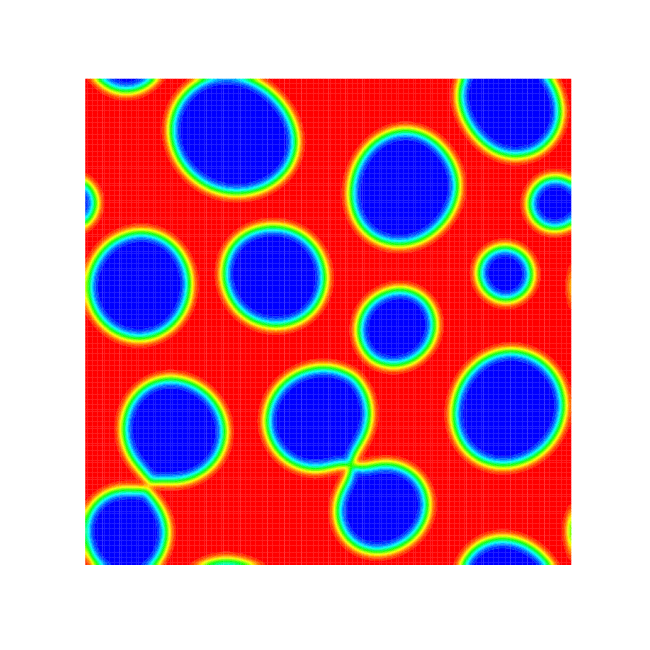}
\label{FIG:step250K}
}
\subfigure[$t^{*} = 625$]
{
\includegraphics[width=0.48 \columnwidth]{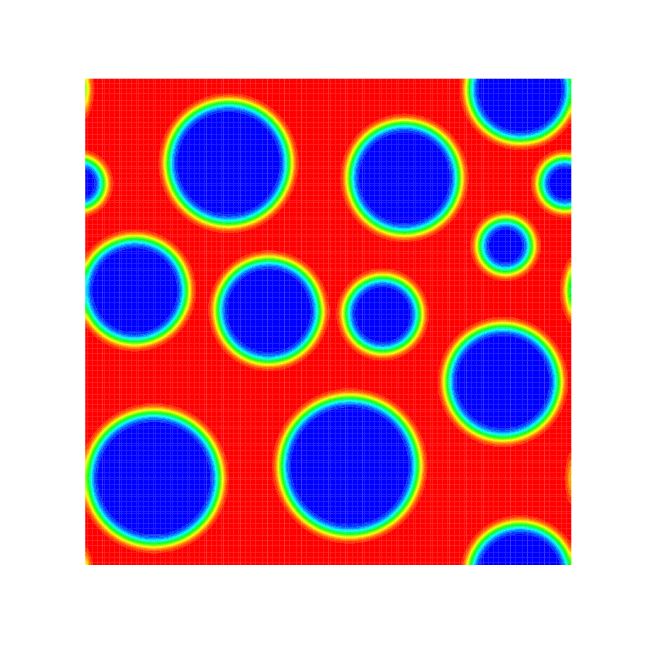}
\label{FIG:step500K}
}
\caption{Contours of density distribution at various moments in the process of phase separation}
\label{FIG:Evolution}
\end{figure}
To investigate the conservation of mass, we also record the mass variation in terms of the whole domain during the process of phase separation. Fig. \ref{FIG:mass variation} illustrates that mass generation and diffusion appears until the system reaches a relatively stable condition where no intensive coalescence or condensation take place. Although the relative value of mass variation is small, it may induce unphysical behaviors at some special conditions\cite{Yang2016Lattice}. For problems referring to continuous separation of different phases, the quasi-incompressible model in literature\cite{Zhang2018} is more reliable.
\begin{figure}[H]
\centering{}
\includegraphics[width=0.48 \columnwidth]{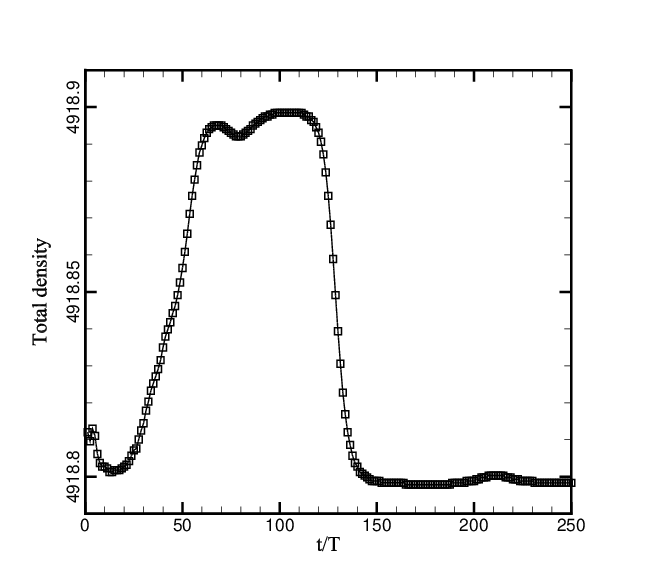}
\caption{Mass variation along with the process of phase separation}
\label{FIG:mass variation}
\end{figure}
\subsection{Rayleigh-Taylor instability}
For the last case, the benchmark problem of Rayleigh-Taylor instability (RTI) is conducted. The RTI is a common and important phenomenon in nature, which occurs when a heavy phase is on top of a light phase with initial perturbation in the interface. This benchmark has been extensively studied by several newly developed numerical approaches\cite{Ding2007,Zu2013Phase,Li2012Additional,Liang2014Phase,Zheng2015Lattice,Wang2015Multiphase,Zhang2018} in order to validate their reliability. The computational domain is a rectangular box with $L_0$$\times$$4L_0$ cells. The initial interface is located at $H(x,y) = 2L_0+0.1L_0\text{cos}(2{\pi}x/L_0)$ and the initial order profile is defined as
\begin{equation}
\phi(x,y)=\frac{\phi_H+\phi_L}{2}+\frac{\phi_H-\phi_L}{2}\text{tanh}\frac{2[y-H(x,y)]}{W}.
\end{equation}
Two dimensionless numbers, Atwood number and Reynolds number, are introduced to characterize RTI and their expressions are defined as follows
\begin{equation}
At = (\rho_H-\rho_L)/(\rho_H+\rho_L),\ Re = \rho_H{L_0}\sqrt{\vert{gL}\vert}/\mu .
\end{equation}
The physical parameters are fixed as $L_0 = 256,\sqrt{gL_0}=0.04,W=5$ and $\sigma=5{\times}10^{-5}$ and the reference time is set to be $T = \sqrt{L_0/{g}At}/{\Delta{t}}$. The CFL number is adjusted to 0.25 to reduce the temporal dissipation. The non-slip boundary condition are applied to the top and bottom sides and periodic boundary condition are implemented to the left and right sides.

\begin{figure}[H]
\centering{}
\subfigure[$t = 1.0T$]
{
\includegraphics[width=0.15 \columnwidth]{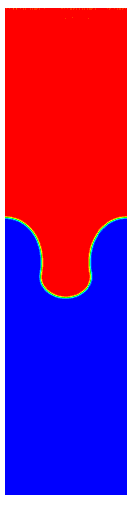}
\label{FIG:cutRe3000At0_5t3_5T}
}
\subfigure[$t = 1.5T$]
{
\includegraphics[width=0.15 \columnwidth]{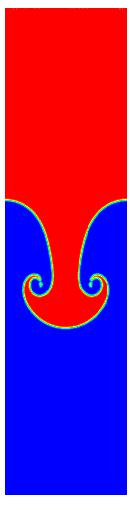}
\label{FIG:cutRe3000At0_5t4_0T}
}
\subfigure[$t = 2.0T$]
{
\includegraphics[width=0.15 \columnwidth]{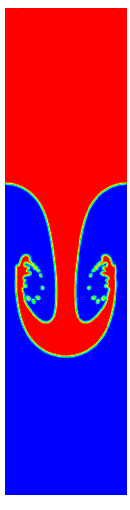}
\label{FIG:cutRe3000At0_5t2_0T}
}
\subfigure[$t = 2.5T$]
{
\includegraphics[width=0.15 \columnwidth]{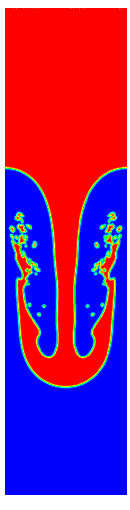}
\label{FIG:cutRe3000At0_5t2_5T}
}
\subfigure[$t = 3.0T$]
{
\includegraphics[width=0.15 \columnwidth]{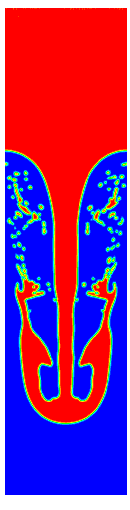}
\label{FIG:cutRe3000At0_5t3_0T}
}
\caption{Time evolution of interface patterns of Rayleigh-Taylor instability at $At = 0.5, Re = 3000$.}
\label{FIG:cutRe3000At0_5}
\end{figure}

To make a comparison between the results obtained by current method and other available data presented in literature\cite{Ding2007,Li2012Additional,Ren2016Improved,Zhang2018}, we first carried out a simulation at the condition of At = 0.5, Re = 3000. Five stages in the evolution of interface are illustrated in Fig. \ref{FIG:cutRe3000At0_5}. As is observed in previous work, the heavy fluid falls down symmetrically by gravity and the light fluid is driven to rise up on the opposite side. The flow patterns at early stages show same characteristics as those results presented in literature\cite{Ren2016Improved,Fakhari2017Improved} since the elongation of interface is still small. As it evolved, breakups near the rolling-up tails of the interface can be observed clearly. The results obtained with current method go through more severe breakups, which is mainly caused by the larger dissipation feature of DUGKS compared to LB method. The interface at the top of flow domain is no longer distinguishable when it comes to the end stage of evolution. Actually, if we make a close inspection of the interfacial differences between the results achieved by A-C equation\cite{Ren2016Improved,Fakhari2017Improved} and that get from C-H equation\cite{Li2012Additional,Zu2013Phase,Liang2014Phase} in the framework of LB theory, it can be found that in the results presented by Allen-Cahn equation the rolling-up tails of the interface tends to break up at an early stage while an elaborated contour of interface rolling-up can be observed and tail breakups are delayed in the results of C-H equation. The rolling-up of interface shown in current case shares some similarities with the interface elongation problem in Sec. \ref{SEC:INTERFACE-CAPTURING TESTS}.\ref{Subsec:elongationinterface}. Both of them undergoes a interface elongation process during which a smoothed interface is stretched and prolongated. As is depicted in Sec. \ref{SEC:INTERFACE-CAPTURING TESTS}, the C-H equation shows a better performance than the conservative A-C equation in the interface elongation test. Hence, it is reasonable to get a more distinguishable interface contour with the C-H equation. Solving the A-C equation in the framework of DUGKS has made its weak points more obvious.

\begin{figure}[H]
\centering{}
\subfigure[bubble front]
{
\includegraphics[width=0.48 \columnwidth]{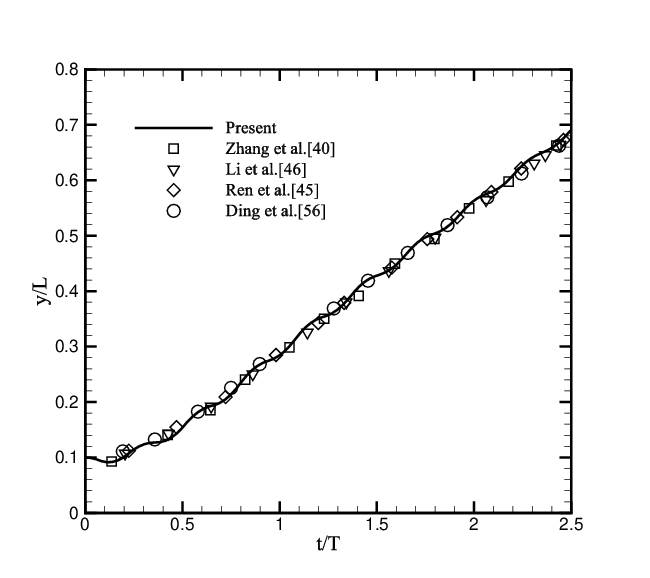}
\label{FIG:falling}
}
\subfigure[spike tip]
{
\includegraphics[width=0.48 \columnwidth]{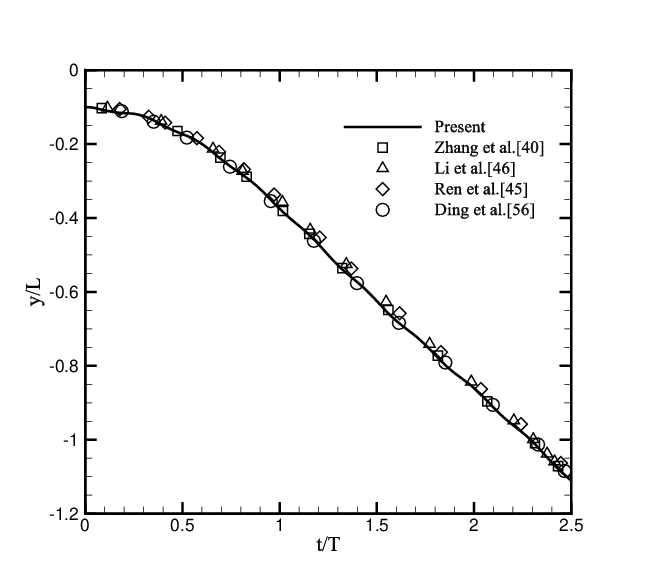}
\label{FIG:rising}
}
\caption{Time evolution of bubble front and spike tip positions. Comparison with the results of Zhang et al\cite{Zhang2018}, Li et al\cite{Li2012Additional}, Ren et al\cite{Ren2016Improved},and Ding et al\cite{Ding2007}.}
\label{FIG:RTIFallingRisingPosition}
\end{figure}

\begin{figure}[H]
\centering{}
\subfigure[$t = 2.0T$]
{
\includegraphics[width=0.15 \columnwidth]{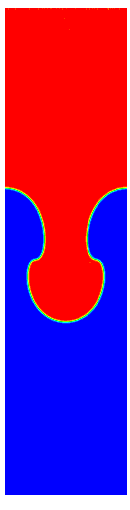}
\label{FIG:cutRe150At0_1t2_0T}
}
\subfigure[$t = 2.5T$]
{
\includegraphics[width=0.15 \columnwidth]{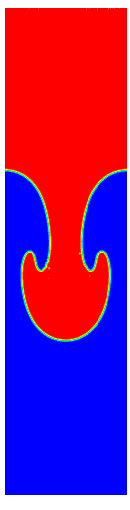}
\label{FIG:cutRe150At0_1t2_5T}
}
\subfigure[$t = 3.0T$]
{
\includegraphics[width=0.15 \columnwidth]{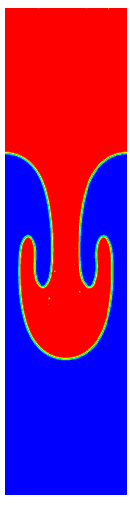}
\label{FIG:cutRe150At0_1t3_0T}
}
\subfigure[$t = 3.5T$]
{
\includegraphics[width=0.15 \columnwidth]{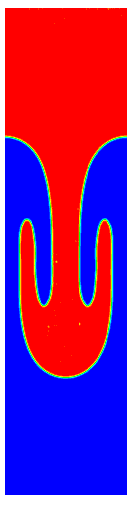}
\label{FIG:cutRe150At0_1t3_5T}
}
\subfigure[$t = 4.0T$]
{
\includegraphics[width=0.15 \columnwidth]{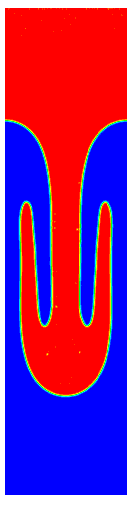}
\label{FIG:cutRe150At0_1t4_0T}
}
\caption{Time evolution of interface patterns of Rayleigh-Taylor instability at $At = 0.1, Re = 150$.}
\label{FIG:cutRe150At0_1}
\end{figure}

\begin{figure}[H]
\centering{}
\subfigure[$t = 2.0T$]
{
\includegraphics[width=0.15 \columnwidth]{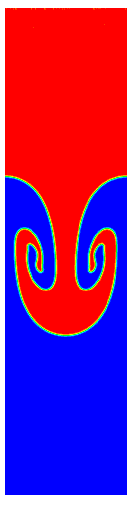}
\label{FIG:cutRe3000At0_1t2_0T}
}
\subfigure[$t = 2.5T$]
{
\includegraphics[width=0.15 \columnwidth]{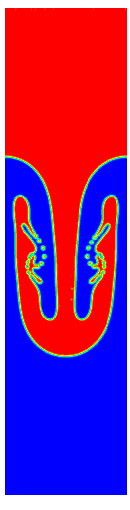}
\label{FIG:cutRe3000At0_1t2_5T}
}
\subfigure[$t = 3.0T$]
{
\includegraphics[width=0.15 \columnwidth]{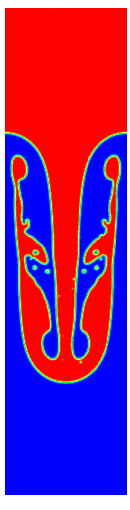}
\label{FIG:cutRe3000At0_1t3_0T}
}
\subfigure[$t = 3.5T$]
{
\includegraphics[width=0.15 \columnwidth]{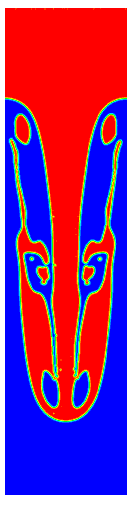}
\label{FIG:cutRe3000At0_1t3_5T}
}
\subfigure[$t = 4.0T$]
{
\includegraphics[width=0.15 \columnwidth]{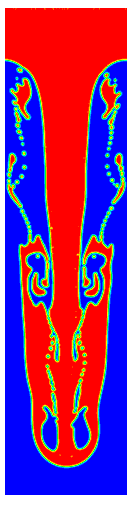}
\label{FIG:cutRe3000At0_1t4_0T}
}
\caption{Time evolution of interface patterns of Rayleigh-Taylor instability at $At = 0.1, Re = 3000$.}
\label{FIG:cutRe3000At0_1}
\end{figure}

Variations in the dimensionless positions of bubble front of the light phase and spike tip of the heavy phase are shown in Fig. \ref{FIG:RTIFallingRisingPosition}. It can be seen that the results presented by current method are in good agreement with the previous works\cite{Zhang2018,Li2012Additional,Ren2016Improved,Ding2007}.
To make a further exploration of the capability of present method, simulations of RTI at two more conditions of $At = 0.1, Re = 150$ and $At = 0.1, Re = 3000$ are carried out and the evolution of interface at five different stages are presented in Fig. \ref{FIG:cutRe150At0_1} and \ref{FIG:cutRe3000At0_1}. At low Reynolds number ($Re = 150$), there are not much differences between the results of Zhang and ours as the interface pattern is simple and no complicated structure is evolved. As the Reynolds number goes higher (Re=3000), large disparities in the evolution of interface patterns can be observed between these two models. The rolling-up tips of the interface break up into small bubbles or drops, which are illustrated by the discrete blue or red spots in Fig. \ref{FIG:cutRe3000At0_1} at $t=2.5T$. With the evolution of the interface, coalescence of these scattered elements can be observed, which means that process of phase separation happens locally. As the development of system, the tails grow thinner as well as longer. At a certain moment, breakups of this slim tails take place and the interface pattern at $t = 4.0T$ shows up. Compared to the evolution process depicted by Zhang\cite{Zhang2018}, the present model fails to give a distinct depiction of interface patterns at a later stage.

\begin{figure}[H]
\centering{}
\subfigure[$t = 2.0T$]
{
\includegraphics[width=0.15 \columnwidth]{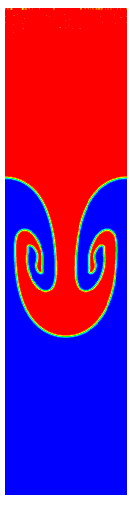}
\label{FIG:cutRe3000At0_1t2_0THCZ}
}
\subfigure[$t = 2.5T$]
{
\includegraphics[width=0.15 \columnwidth]{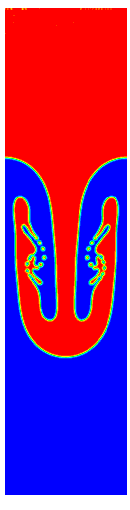}
\label{FIG:cutRe3000At0_1t2_5THCZ}
}
\subfigure[$t = 3.0T$]
{
\includegraphics[width=0.15 \columnwidth]{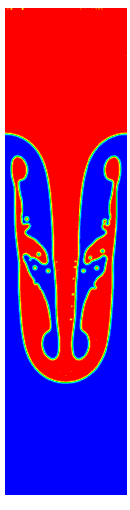}
\label{FIG:cutRe3000At0_1t3_0THCZ}
}
\subfigure[$t = 3.5T$]
{
\includegraphics[width=0.15 \columnwidth]{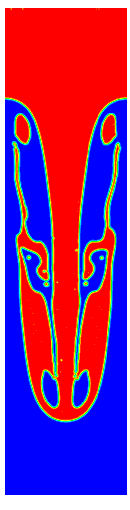}
\label{FIG:cutRe3000At0_1t3_5THCZ}
}
\subfigure[$t = 4.0T$]
{
\includegraphics[width=0.15 \columnwidth]{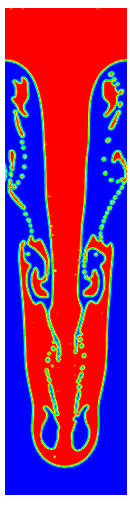}
\label{FIG:cutRe3000At0_1t4_0THCZ}
}
\caption{Time evolution of interface patterns of Rayleigh-Taylor instability by HCZ model at $At = 0.1, Re = 3000$.}
\label{FIG:cutRe3000At0_1HCZ}
\end{figure}

To investigate the influence of different hydrodynamic models, another case using the hydrodynamic part of HCZ model\cite{HE1999642} is conducted. Both Fakhari\cite{Fakhari2016A} and Zhang\cite{Zhang2018} adopts this kind of model to solve the mass and momentum equations. All of the parameters were kept the same as the above one with a condition of $At = 0.1, Re = 3000$. The time evolution of interface pattern is shown in Fig \ref{FIG:cutRe3000At0_1HCZ}. At early stages, the interface patterns obtained by two different models are almost identical. Although slight disparities can be observed at the last two stages, the overall flow patterns obtained by HCZ model are nearly the same as the results shown in Fig. \ref{FIG:cutRe3000At0_1}. Hence, the influence of different hydrodynamic models can be neglected.
With the comparisons conducted in this subsection, we can conclude that the failure of DUGKS in the detailed depiction of interface during the evolution of RTI is to blame for not only the deficiencies of A-C equation, which were ignored by Wang\cite{Wang2016Comparative}, but also the relatively larger dissipations of DUGKS compared to the LB method. To overcome this problem, a high-order scheme of DUGKS\cite{Wu2018Third} or adaptive mesh refinement technology needs to be implemented.

\section{CONCLUSION}\label{sec:CONCLUSION}
In this article, a phase-field based model for the simulation of two-phase flows is developed in the framework of DUGKS. The conservative Allen-Cahn equation is used to capture the interface and incompressible hydrodynamic models are employed to solve the velocity and pressure field. The macroscopic equations can be recover exactly from the discrete kinetic models through the Chapmann-Enskog analysis.

The performance of proposed model are validated thoroughly by a series of numerical tests. In the interface-capturing tests, our model presents reliable results which are in good agreement with LB method for the convergence rate and numerical dispersion at various P\`{e}clet numbers and mobility coefficients. A key deficiency of Allen-Cahn equation, ignored by Wang\cite{Wang2016Comparative}, is discovered in the simulation of interface elongation test. It has been verified that this drawback has a great impact on the results of Rayleigh-Taylor instability. In the binary flow tests, stationary bubbles with various density ratios are investigated. The Laplace's law has been verified and relations between the maximum magnitude of spurious velocity and the Laplace number at various density ratios are disclosed quantitatively. Compared to previous work, a lower magnitude of spurious velocity can be achieved with the present method. In the case of layered Poiseuille flow, the numerical results of velocity profile obtained at high viscosity ratios agree well with the analytical solution. The comparison between two different schemes of viscosity calculation is conducted and similar phenomena to that shown by Zu and He\cite{Zu2013Phase} are observed. The capability of present model in the simulations involving high density ratio is further validated by the spinodal decomposition test. The phenomenon of mass generation or consummation is observed and the parameter that conservativeness qualifies has been emphasized. The present model fails to give a detailed depiction of the interface patterns in the evolution process of Rayleigh-Taylor instability, which is mainly due to the numerical dissipation of DUGKS. To overcome this problem, a high-order scheme or adaptive mesh refinement technique needs to be implemented.

In conclusion, we makes a preliminary research on the performance of the Allen-Cahn based two-phase model under the framework of DUGKS. Satisfying results are obtained in several benchmark tests. Compared to Zhang's model\cite{Zhang2018}, a high density ratio of 1000 can be achieved and calculation of source terms referring to high-order derivatives has been simplified. Capturing of the subtle interfaces during the evolution of Rayleigh-Taylor instability is failed. High-order scheme needs to be developed in future research.

\section*{Acknowledgements}
The project has been financially supported by the National Natural Science Foundation of China (Grant No. 11472219), the 111 Project of China (B17037), and the ATCFD Project (2015-F-016).

\appendix*
\section{Chapman-Enskog analysis}
In this section, the macroscopic equations are recovered from the discrete kinetic equation with the application of Chapman-Enskog expansion. With the introduction of a small parameter $\epsilon$, the discreted distribution function and derivative operators in Eq. (\ref{DKE}) can be expanded as
\begin{subequations}
\begin{equation}
\psi_i = \psi_i^{(0)} + \epsilon\psi_i^{(1)}+\epsilon^2\psi_i^{(2)}+\cdots,
\end{equation}
\begin{equation}
\partial_t = \epsilon\partial_{t_0}+\epsilon^2\partial_{t_1},\nabla = \epsilon\nabla_0,S_i=\epsilon{S_i}^{(0)}.
\end{equation}
\end{subequations}
Substitute the above equation into Eq. (\ref{DKE}) and rearrange each item based on the power of $\epsilon$, we have
\begin{subequations}
\begin{align}
\epsilon^0 &:\psi_i^{(0)} = \psi_i^{(eq)},
\label{Eq:CE-a}
\\
\epsilon^1 &:\partial_{t_0}\psi_i^{(0)}+\bm{\xi}_i\cdot\nabla_0{\psi_i^{(0)}}=-\frac{1}{\tau}\psi_i^{(1)}+S_i^{(0)},
\label{Eq:CE-b}
\\
\epsilon^2 &:\partial_{t_0}\psi_i^{(1)}+\partial_{t_1}\psi_i^{(0)}+\bm{\xi}_i\cdot\nabla_0{\psi_i^{(1)}}=-\frac{1}{\tau}\psi_i^{(2)}.
\label{Eq:CE-c}
\end{align}
\label{Eq:C-E}
\end{subequations}
First we give a detailed derivation of the A-C equation Eq. (\ref{Eq:interfaceEq}).

The moments of $f_i$ and its corresponding source terms can be calculated by Eq. (\ref{Eq:phiEq}) and Eq. (\ref{phiSource}), i.e.,
\begin{equation}
\sum_if_i^{eq}=\phi,\sum_i\bm{\xi}_if_i^{eq}=\phi\bm{u},\sum_i\bm{\xi}_i\bm{\xi}_if_i^{eq}=\phi{RT}\bm{I},
\end{equation}
\begin{equation}
\sum_i{F^{(0)}_i}=0,\sum_i\bm{\xi}_i{F^{(0)}_i}=\bm{F}_\phi^{(0)}=\partial_{t_0}(\phi\bm{u})+\epsilon\partial_{t_1}(\phi\bm{u})+\theta{RT}\frac{\nabla_0\phi}{\vert\nabla\phi\vert}.
\end{equation}
Replacing $\psi_i$ with $f_i$ in Eq. (\ref{Eq:C-E}) and taking the zeroth- and first-order moments of Eq. (\ref{Eq:CE-b}), we have
\begin{subequations}
\begin{align}
\partial_{t_0}\phi+\nabla_0(\phi\bm{u}) &= 0,\label{Eq:CEphi-a}\\
\partial_{t_0}(\phi\bm{u})+RT\nabla_0\phi &= -\frac{1}{\tau_f}M^{(1)}+\bm{F}_{\phi}^{(0)}.\label{Eq:CEphi-b}
\end{align}
\end{subequations}
The zeroth-order moment of Eq. (\ref{Eq:CE-c}) is given as
\begin{equation}
\partial_{t_1}\phi+\nabla_0{M^{(1)}}=0.
\label{Eq:CEphi-c}
\end{equation}
Calculate $M^{(1)}$ in Eq. (\ref{Eq:CEphi-b}) and substitute it into Eq. (\ref{Eq:CEphi-c}), and then we have
\begin{equation}
\partial_{t_1}\phi =\tau_f{RT}\nabla_0\Big(\nabla_0\phi-\theta\frac{\nabla_0\phi}{\vert\nabla\phi\vert}-\epsilon\partial_{t_1}(\phi\bm{u})\Big).
\label{Eq:Phi2nd}
\end{equation}
Combining Eq. (\ref{Eq:Phi2nd}) with Eq. (\ref{Eq:CEphi-a}) and neglecting the term of $O(\epsilon^3)$, the final A-C equation Eq. (\ref{Eq:interfaceEq}) can be exactly recovered with $M_\phi=\tau_f{RT}$.

Next the recovery of hydrodynamic equations are explained with elaboration.

The moments of $g_i$ and its corresponding source terms can be computed from Eq. (\ref{Eq:HydroEq}) and Eq. (\ref{HydroSource}), i.e.,
\begin{equation}
\begin{aligned}
&\sum_i{g_i^{(eq)}}=0, \sum_i{\bm{\xi}_ig_i^{(eq)}}=\rho\bm{u},\sum_i{\bm{\xi}_i\bm{\xi}_ig_i^{(eq)}}=p\bm{I}+\rho\bm{u}\bm{u},\\
&\sum_i{\bm{\xi}_i\bm{\xi}_i\bm{\xi}_ig_i^{(eq)}}=RT\rho\tilde{3}\bm{uI}=RT\rho(\delta_{\alpha\beta}u_{\gamma}+\delta_{\beta\gamma}u_{\alpha}+\delta_{\gamma\alpha}u_{\beta}),
\end{aligned}
\end{equation}
\begin{equation}
\begin{aligned}
\sum_i{G_i} &= \bm{u}\cdot\nabla{\rho},\sum_i\bm{\xi}_i{G_i}=\bm{F}_s+\bm{G}=\bm{F},\\
\sum_i\bm{\xi}_i\bm{\xi}_i{G_i} &= [\bm{uF}+\bm{Fu}]+RT[\bm{u}\nabla\rho+\nabla\rho\bm{u}+(\bm{u}\cdot\nabla\rho)\bm{I}]\\
&=[u_\alpha{F_\beta}+u_\beta{F_\alpha}]+RT[u_\alpha{\partial_{\beta}\rho}+u_\beta{\partial_{\alpha}\rho}+u_{\gamma}\partial_{\gamma}\rho\delta_{\alpha\beta}].
\end{aligned}
\end{equation}
Since we have the following relations in terms of the conservative variables:
\begin{equation}
\sum_ig_i=0,\sum_i\bm{\xi}_ig_i=\rho\bm{u}.
\end{equation}
It is easy to get
\begin{equation}
\sum_ig_i^{(k)}=0,\sum_i\bm{\xi}_ig_i^{(k)}=0,k>0.
\label{Eq:HydroNeqMoments}
\end{equation}
Replacing $\psi_i$ with $g_i$ in Eq. (\ref{Eq:C-E}) and taking the zeroth- and first-order moments of Eq. (\ref{Eq:CE-b}), we have
\begin{subequations}
\begin{align}
\nabla\cdot\bm{u} &= 0,\label{Eq:ico}\\
\partial_{t_0}(\rho\bm{u})+\nabla_0(\rho\bm{uu}+p\bm{I}) &= -\frac{1}{\tau_g}\Pi^{(1)}+\bm{F}^{(0)},\label{Eq:Euler}
\end{align}
\label{Eq:HydroDKE}
\end{subequations}
where $\Pi^{(1)}=\sum_i\xi_i{g_i}^{(1)}=0$.

The zeroth- and first-order moments of Eq. (\ref{Eq:CE-c}) are expressed as
\begin{subequations}
\begin{align}
\nabla_0\Pi^{(1)} &= 0,\\
\partial_{t_1}(\rho\bm{u}) &= -\nabla_0(\sum_i\bm{\xi}_i\bm{\xi}_i{g}_i^{(1)}),
\label{Eq:HydroShear}
\end{align}
\end{subequations}
where
\begin{equation}
\begin{aligned}
\sum_i\xi_{i\alpha}\xi_{i\beta}{g}_i^{(1)} &= -\tau_g\Big[\partial_{t_0}({\rho{u}_\alpha{u_\beta}+p\delta_{\alpha\beta}})
+RT[\nabla_{0\alpha}(\rho{u}_\beta)+\nabla_{0\beta}(\rho{u}_\alpha)+\nabla_{0\gamma}(\rho{u}_\gamma)]-\sum_i\xi_{i\alpha}\xi_{i\beta}G_i^{(0)}\Big]\\
&=-\tau_gRT\Big[\rho\partial_{0\alpha}u_\beta+\rho\partial_{0\beta}u_\alpha\Big]+O(u^3).
\end{aligned}
\end{equation}
Combining Eq. (\ref{Eq:HydroShear}) with Eq. (\ref{Eq:Euler}), we get the momentum equation in final form:
\begin{equation}
\partial_t(\rho\bm{u})+\nabla\cdot(\rho\bm{uu}+p\bm{I})=\nabla\cdot[\rho\nu(\nabla\bm{u}+\nabla\bm{u}^T)]+F,
\end{equation}
where $\nu = \tau_gRT$.

Since the computation of dynamic pressure in Eq. (\ref{Eq:dynamicPressure}) is a bit complicated, a detailed derivation is given below.
The zeroth-order moment of  Eq. (\ref{Eq:Tilde}) with $\psi_i$ replaced by $g_i$ is given as
\begin{equation}
\sum_i\tilde{g_i}=\frac{2\tau_g+\Delta{t}}{2\tau_g}\sum_i{g_i}-\frac{\Delta{t}}{2\tau_g}\sum_i{g_i^{eq}}-\frac{\Delta{t}}{2}\bm{u}\cdot\nabla\rho,
\end{equation}
where $g_i$ can be divided into the equilibrium part  $g_i^{eq}$ and non-equilibrium part $g_i^{neq}$. The the above equation can be rearranged as
\begin{equation}
\sum_i\tilde{g_i}+\frac{\Delta{t}}{2}\bm{u}\cdot\nabla\rho = \sum_i{g_i^{eq}}+\frac{2\tau_g+\Delta{t}}{2\tau_g}\sum_i{g_i^{neq}} = 0.
\end{equation}
Subtracting $\tilde{g}_0$ from the left hand side of the above equation, we get
\begin{equation}
\sum_{i\neq{0}}\tilde{g_i}+\frac{\Delta{t}}{2}\bm{u}\cdot\nabla\rho= \sum_{i\neq{0}}{g_i^{eq}}+\frac{2\tau_g+\Delta{t}}{2\tau_g}\sum_{i\neq{0}}{g_i^{neq}}-\underbrace{\frac{\Delta{t}}{2}G_0}_{O(u^3)}.
\end{equation}
With the help of Eq. (\ref{Eq:HydroNeqMoments}), the above equation can be rewritten as
\begin{equation}
\sum_{i\neq{0}}\tilde{g_i}+\frac{\Delta{t}}{2}\bm{u}\cdot\nabla\rho= -{g_0^{eq}}-\frac{2\tau_g+\Delta{t}}{2\tau_g}{g_0^{neq}}.
\end{equation}
The non-equilibrium term ${g_0^{neq}}$ can be dropped since its value is tiny compared to that of ${g_0^{eq}}$. At last we have
\begin{equation}
\sum_{i\neq{0}}\tilde{g_i}+\frac{\Delta{t}}{2}\bm{u}\cdot\nabla\rho = -{g_0^{eq}}.
\end{equation}
The dynamic pressure can be finally calculated by Eq. (\ref{Eq:HydroEq}).

\bibliography{PREYang}
\bibliographystyle{apsrev4-1}

\end{document}